\documentclass[letterpaper,usenatbib,useAMS,twocolumn]{mn2e}
\usepackage{color}
\usepackage{float} 
\usepackage[fleqn]{amsmath}
\usepackage{epsfig,floatflt}
\usepackage{natbib}
\usepackage{subfigure}
\usepackage{amssymb}
\usepackage{multirow}
\usepackage{ulem}
\usepackage{bm}
\usepackage{url}

\DeclareMathSymbol{\la}{3}{AMSa}{46}
\DeclareMathSymbol{\ga}{3}{AMSa}{38}
\newcommand{\LCDM}{ $\Lambda$CDM }
\newcommand{\be}{\begin{equation}} \newcommand{\ee}{\end{equation}}
\newcommand{\ba}{\begin{eqnarray}} \newcommand{\ea}{\end{eqnarray}}
\newcommand{\brr}{\begin{array}} \newcommand{\err}{\end{array}}

\newcommand{\kmsMpc}{km s$^{-1}$ Mpc$^{-1}$}
\newcommand{\Mpch}{$h^{-1}$Mpc}
\newcommand{\VolMpch}{$h^{-3}$Mpc$^{3}$}
\newcommand{\sourcePsi}{{S_\Psi}}

\begin{document}

\title{Precision cosmology with voids: definition, methods, dynamics}

\author[G. Lavaux \& B.~D. Wandelt]{Guilhem Lavaux$^{1,2}$, Benjamin~D. Wandelt$^{1,2}$\\
$^{1}$ Department of Physics, University of Illinois at Urbana-Champaign, 1110 West Green Street, Urbana, IL 61801-3080, USA\\
$^{2}$ California Institute of Technology, Pasadena, CA 91125, USA\\
}

\maketitle

\begin{abstract}
  We propose a new definition of cosmic voids based on methods of
  Lagrangian orbit reconstruction as well as an algorithm to find them
  in actual data called DIVA. Our technique is intended to
  yield results which can be modeled sufficiently accurately to create
  a new probe of precision cosmology.
  We then develop an analytical model of the ellipticity of voids 
  found by our method
  based on Zel'dovich approximation. We measure in $N$-body simulation
  that this model is precise at the $\sim$0.1\% level for the mean
  ellipticity of voids of size greater than $\sim$4\Mpch.  We estimate
  that at this scale, we are able to predict the ellipticity with an
  accuracy of $\sigma_\varepsilon \sim 0.02$.  Finally, we compare the
  distribution of void shapes in $N$-body simulation for two different
  equations of state $w$ of the dark energy. We conclude that our
  method is far more accurate than Eulerian methods and is therefore
  promising as a precision probe of dark energy phenomenology.
\end{abstract}

\section{Introduction}
\label{sec:intro}

Large empty regions of space, called voids, represent the majority of
the volume of the present Universe. They were first discovered in
observations by \cite{GT78}, \cite{Joeveer78} and
  \cite{TullyFisher78}, followed later by \cite{Kirshner81} and more
largely in the CfA redshift catalogue \citep{Lapparent86}. This discovery was followed
by a large amount of theoretical work. The first gravitational instability model was
given by \cite{HoffmanShaham82}, quickly followed by \cite{HSW83} for an infinite
size regular mesh of void and by \cite{HOR83} for the impact of cosmology on their
evolution. Other work studied the general self-similar evolution of voids
in Einstein-de-Sitter universes 
\citep{Bertschinger83,Bertschinger85}.

However, as the voids are intrinsically large and the surveys at that
time were small, we only detected a small number of them. This has
hindered their use as a cosmological probe for a long time
[except some constraints on their maximal size compatible
with CMB observations by e.g. \cite{Blumenthal92}].  This situation has changed with the advent
of deep and wide galaxy surveys such as the Sloan Digital sky survey
\citep[SDSS, ][]{SDSS}, 2dFGRS \citep{TwoDF}, 2MRS \citep{TwoMRS} and
now the 6dFGS \citep{SixDF}.  Still we miss a clear and simple
definition of voids that would allow us to use them as a precision
cosmological probe. In this paper, we investigate, analytically and numerically using $N$-body simulations, a new algorithm for finding voids
in Large Scale structure surveys and an analytical model that accurately
predicts the properties of voids found by this method as a function
of cosmology.

During the last decade, several algorithms to find voids have been built. They
are separated in three broad classes. In the first class, the void finders try to find
regions empty of galaxies \citep{Kauffman91,ElAd97,HoyleVogeley2002,Patiri06,FosterNelson09}.
The second class of void finders try to identify voids as geometrical structures
in the dark matter distribution traced by galaxies
\citep{PlionisBasilakos02,Colberg05,Shandarin06,wvf,zobov}.
The third class identifies structures dynamically by checking
gravitationally unstable points in the distribution of dark matter
\citep{Hahn07,VoidsGravity08}. At the same time, $N$-body simulations focused
on the studies of voids in a cosmological context were flourishing \citep{Martel90,Regos91,vdW93,GLKH03,Benson03,Colberg05}.
Recently, \cite{AspenAmsterdam}
made a comparison which shows that, even if these currently available void finder techniques find
approximately the same voids, the details of the shapes and sizes found by each of the void finders may be
significatively different.  This problem is further enhanced by the existence of
ad-hoc parameters in most of the existing void finders, which changes the exact
definition of voids and does not allow reliable cosmological
predictions. One aspect of voids that is also often left aside is the
hierarchical structures of voids. So far, apart from ZOBOV
\citep{zobov} and the related Watershed Void Finder (WVF) method \citep{wvf}, which are parameter free,
no void finder tries to identify correctly the hierarchy of
voids-in-voids and clouds-in-voids \citep{ShethWeygaert04}.  Another problem of these void
finders comes from their Eulerian nature: they try to find structures
that are not necessarily in the same dynamical regime (linear or
non-linear), which complicates the building of an analytical model.

We propose studying a new void finder that belongs to the third class
of these void finders. It is based on the success of both the
Monge-Amp\`ere-Kantorovitch (MAK) reconstruction of the orbits of
galaxies \citep{Brenier2002,moh2005,lavaux08} and the Zel'dovich
approximation \citep{zeldov70}. This method is based on finding a way
to compute the Lagrangian coordinates of the objects at their present
position. The study of voids in Lagrangian coordinates is
  not new.  The evolution of voids in the adhesion approximation has
  been studied by \cite{SahniShandarin94} to understand the
  formation and evolution of voids and their inner substructure in
  a cosmological context.
  Later, \cite{SahniShandarin96} emphasized the precision of the Zel'dovich
  approximation for studying void dynamics compared to higher order
  perturbation theory, either Lagrangian or Eulerian. However, no void finder method has yet
tried to take advantage of the Zel'dovich approximation for detecting and studying voids in real data. The voids detected with this method are going to be intrinsically different than the one found using standard Eulerian void finder. This hardens the possibility of making a void-by-void comparison of the different methods.

The use of Lagrangian coordinates gives one immediate advantages compared 
to standard void finding: the Lagrangian displacement field is still largely in
the linear regime even at $z=0$ and especially for voids. This allows
us for the first time to make nearly exact analytical computation on
the dynamical and geometrical properties of voids in Large scale
structures. The MAK reconstruction is thus particularly adapted to
study the dynamics of voids. However, there is an apparent cost to
pay: we lose the intuitive way of defining voids as ``holes'' in a
distribution of galaxy, that is the place where matter is not
anymore. On the other hand, we gain the physical understanding that
voids correspond to regions from which matter is escaping.

The dynamics of voids may provide a wealth of information on dark
energy without the need for any new survey. The first obvious probe
 of dark energy
properties comes from the study of the linear growth factor. Its
evolution with redshift depends, among the other cosmological
parameters, on the equation of state $w$ of the Dark Energy. In this
work, we assume that $w$ is independent of the redshift. We note that in galaxy
surveys, our method is going to be sensitive to bias but not more
than the direct approach to void finding. Indeed, void finders of the first class are
sensitive to the selection function of galaxies. Generally this is
done by limiting the survey to galaxies with an apparent magnitude
below some designated threshold. Changing this selection function of
the galaxies acts on the boundaries of the detected voids, which thus
changes the geometry of these voids. From the point of view of void
finders, this will also act as a ``bias''. The method that we propose
has a more conventional dependence on the bias by using the dark
matter distribution inferred from the galaxy distribution. The
advantage is that this bias could be calibrated. One exact calibration
consists in comparing peculiar velocities reconstructed using MAK to
observed velocities \citep{lavaux09}. Additionally, there are a number
of other complementary ways of determining bias from galaxy redshift
surveys
\citep[e.g.][]{benoist96,Norberg01,tegmark04,Erdogdu2005,tegmark06,Percival07}.

This paper is a first of a series studying the properties of voids
found by our void finder. It is organised as follows. First, we recall
the theory of the Monge-Amp\`ere-Kantorovitch reconstruction in
Section~\ref{sec:mak}. Then, we explain how we can use reconstructed
orbits as an alternative way to detect and characterise voids. This
corresponds to the core of DIVA, our void finder through Dynamical
Void Analysis, and is explained in Section~\ref{sec:diva}.  In
Section~\ref{sec:analytic_voids}, we model analytically the voids
found by DIVA. In Section~\ref{sec:nbody_test}, we test our void
finder on $N$-body simulations. We also check our analytical model
against the results of the simulations for two cosmologies. In
Section~\ref{sec:discuss_definitions}, we compare DIVA to earlier
existing void finders. In Section~\ref{sec:conclusion}, we conclude.

\section{The Monge-Ampere-Kantorovitch reconstruction}
\label{sec:mak}

The Monge-Amp\`ere-Kantorovitch reconstruction (MAK) is a method
capable of tracing the trajectories of galaxies back in time
using an approximation of the complete non-linear dynamics.
It is a Lagrangian method, as PIZA \citep{CroftGaztanaga97} or
the Least-Action method \citep{Peebles89}. The MAK reconstruction
is discussed in great detail in \cite{Brenier2002}, \cite{moh2005} and
\cite{lavaux08}. It is based on the hypothesis that, expressed
in comoving Lagrangian coordinates, the displacement field of the
dark matter particles is convex and potential.
Since then, this hypothesis has been  justified by the success of the
method on $N$-body simulations. With the local mass conservation,
this hypothesis leads to the equation of Monge-Amp\`ere:
\begin{equation}
  \text{det}_{i,j} \frac{\partial^2 \Phi}{\partial q_i \partial q_j} =
  \frac{\rho\left({\bf x}({\bf q})\right)}{\rho_0} ,\label{eq:ma}
\end{equation}
with ${\bf q}$ the comoving Lagrangian coordinates, ${\bf x}({\bf q})$
the change of variable between Eulerian (${\bf x}$) and Lagrangian
coordinates (${\bf q}$), ${\rho}({\bf x})$ the Eulerian dark matter density and
${\rho_0}$ the initial comoving density of the Universe, assumed
homogeneous.  \cite{Brenier2002} showed that solving this equation is
equivalent to solving a Monge-Kantorovitch equation, where we seek to
minimise
\begin{equation}
  I\left[{\bf q}({\bf x})\right] = \int_{\bf x}\;\text{d}^3 {\bf x}
  \rho({\bf x}) \left({\bf x} - {\bf q}\left({\bf x}\right)\right)^2,\label{eq:mk}
\end{equation}
according to the change of variable ${\bf q}({\bf  x})$.
Discretising this integral, we obtain
\begin{equation}
  S_\sigma = \sum_i \left({\bf x}_i - {\bf q}_{\sigma(i)}\right)^2 \label{eq:disc_mak},
\end{equation}
with $\sigma$ a permutation of the particles, ${\bf q}_j$ distributed
homogeneously, ${\bf x}_i$ distributed according to the distribution
of dark matter. Doing so, we obtain a discretised version of the
mapping ${\bf q} \rightarrow {\bf x}({\bf q})$ on a grid.

To solve for the problem of minimising Eq.~\eqref{eq:disc_mak} with
respect to $\sigma$, we wrote a high-performance algorithm that has
been parallelised using MPI. This algorithm is based on the Auction
algorithm developed by \cite{Bertsekas79}. It has an overall time
complexity for solving cosmological problems empirically between
$O(n^2)$ and $O(n^3)$ (at worst) with $n$ the number density of mesh
elements $\{ {\bf q}_j\}$.\footnote{An implementation of the algorithm
  is currently available on the first author's website
  http://www.iap.fr/users/lavaux/code.php.}


\section{The Void Finder by Orbit reconstruction}
\label{sec:diva}

In this section, we describe our void finder DIVA (for
  DynamIcal Void Analysis). First, we define in
Section~\ref{sec:definition} what we call a void in this work. Second,
in Section~\ref{sec:ellipticity}, we make use of the displacement
field in the immediate neighbourhood of a void to define the
ellipticity arising from tidal field effects, which we also call {\it tidal ellipticity}. In Section~\ref{sec:euler_epsilon}, we define
the Eulerian ellipticity of our voids. In Section~\ref{sec:smoothing}, we
discuss the impact of smoothing in Lagrangian coordinates to compute 
void properties.

In later sections, we use pure dark
matter $N$-body simulations to check the adequacy of the voids found
using MAK reconstructed displacement field and the one detected in the
simulated displacement field. The results given by the analytical
models are then compared to the one given by the simulated field for
two equations of state of the Dark Energy.

\subsection{Definition of a Void}
\label{sec:definition}

So far, voids have only been described using a purely geometrical
Eulerian approach. Typically, as mentioned in the introduction, 
a void is an empty region delimited by
either sphere or ellipsoid fitting or by using isodensity
contours. We propose here to use a Lagrangian approach and use
the mapping between Lagrangian, ${\bf q}$, and Eulerian coordinates,
${\bf x}$ as a better probe for voids. In the rest of this article, we
will consider these two coordinates to be linked by the displacement
field $\bm{\Psi}$:
\begin{equation}
  {\bf x}({\bf q}) = {\bf q} + \bm{\Psi}({\bf q})\,.
\end{equation}
We now define the source $\sourcePsi$ of the displacement field by
\begin{equation}
  \sourcePsi({\bf q}) =
    \sum_{i=1}^3 \frac{\partial \Psi_i}{\partial q_i}\,. \label{eq:delta}
\end{equation}
As the displacement field is taken to be potential it is strictly
sufficient to only look at $\sourcePsi$ to study ${\bf \Psi}$.

We now define the position of a {\it candidate void} centre by looking
at maxima of $\sourcePsi$ in Lagrangian coordinates.\footnote{This maxima corresponds in terms of the primordial density field to what is sometimes called a protovoid \citep{Blumenthal92,Piran93,Goldwirth95}.}  This will
effectively catch the source of displacement and from where the void
is expanding. The other, practical, advantage is that
$\sourcePsi$ is quite close to the opposite of the linearly extrapolated initial density
perturbations of the considered patch of universe
\citep{moh2005}. So we can use the usual power spectrum to study
most of the statistics of this field. So the main approximation we use
in the rest of this study is that the primordial density field power
spectrum is a good proxy for the power spectrum of the seed of
displacement and that this displacement is a Gaussian random field.

From ${\bf \Psi}$, we define the matrix $T_{l,m}$ of the shear of the displacement, which is linked to the Jacobian matrix:
\begin{eqnarray}
 J_{l,m}({\bf q}) & = & \frac{\partial x_l}{\partial q_m} = \delta_{l,m} +
 \frac{\partial \Psi_l}{\partial q_m}({\bf
   q}) = \delta_{l,m} + T_{l,m}\,, \label{eq:tidal_field} \\
 J({\bf q}) & = & |J_{l,m}|\,,
\end{eqnarray}
with 
\begin{equation}
  T_{l,m}({\bf q}) = \frac{\partial \Psi_l}{\partial q_m}\,. \label{eq:tidal_def}
\end{equation}
$J$ is the Jacobian of the coordinate transformation ${\bf
  q}\rightarrow {\bf x}$. Geometrically, $J$ specifies how an
infinitely small patch of the Universe expanded, in comoving coordinates, from high redshift to
$z=0$. We put $\lambda_i({\bf q})$ the three eigenvalues of
$T_{l,m}({\bf q})$ and sort them such that $\lambda_1({\bf q}) >
\lambda_2({\bf q}) > \lambda_3({\bf q})$. Among the candidate voids,
we select only voids that have strictly expanded, which equivalently
means that $J > 1$. We may now define three classes of voids
that are inspired from the usual classes of observable large scale
structures for galaxies:
\begin{itemize}
\item {\bf true voids} for which $\lambda_1 > 0, \lambda_2 > 0, \lambda_3 >
  0$. These should be the most evident and easily detectable voids as
  they consist in regions which are expanding in the three directions
  of space.
\item {\bf pancake voids} for which $\lambda_1 > 0, \lambda_2 > 0, \lambda_3
  < 0$. The pancake voids are closing along one direction of space but
  expanding along the two other directions. With a geometrical
  analysis, this case cannot be distinguished from the true void
  case. However the dynamical analysis is capable of that, and this
  will cause a crucial difference to the analysis as we will see
  later. In practice they represent a substantial fraction of the
  voids. 
\item {\bf filament voids} for which $\lambda_1 > 0, \lambda_2 < 0,
  \lambda_3 < 0$. 
\end{itemize}
We refer to Section~\ref{sec:analytics_cosmology} for the quantitative relative
number of voids in each of class.
As we will see, the distinction between those cases is important to
quantify the shape and properties of voids that we observe at the
present time.  We discuss our definition of voids, and compare it to other
void finders, in Section~\ref{sec:discuss_definitions}. 


\begin{figure*}
  \includegraphics[width=\hsize]{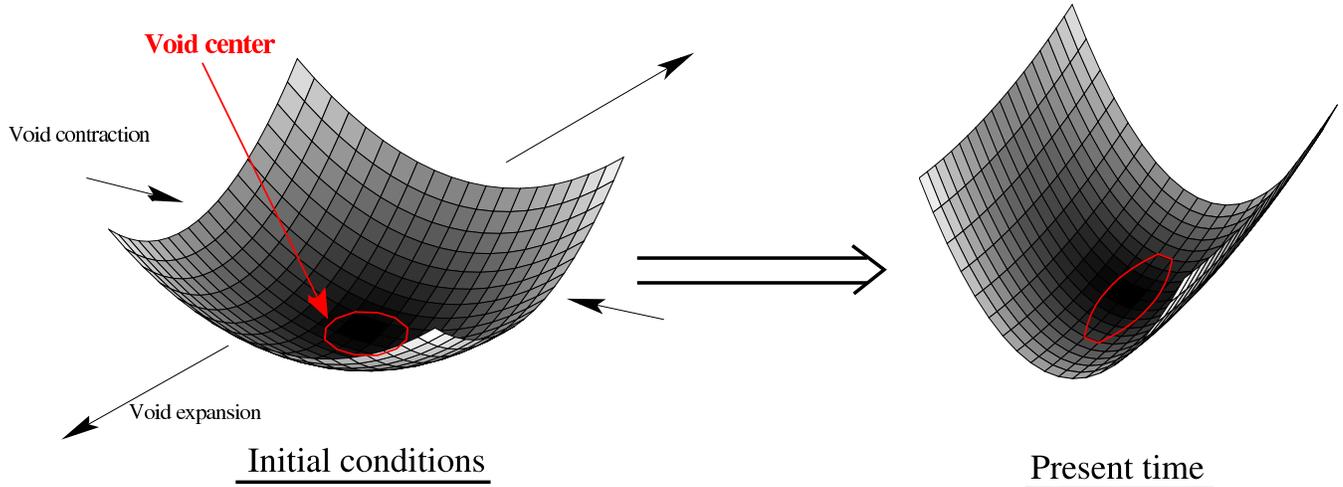}
  \caption{\label{fig:transformations} {\it Picture of a void and the formation
    of intrinsic ellipticity} -- We represent in this figure the 
    central idea of the definition of a void and its ellipticity. We take voids
  as maxima in $\sourcePsi$. They correspond to first order to minima of the primordial density field represented here by painted surface. These minima undergo an overall expansion from initial conditions to present time. The shape of the Void is defined locally at the minimum. The ellipticity is equal to the square root of the ratio of the axes of the ellipsoid which locally fits the surface.}
\end{figure*}

Here, we have not yet touched the issue of defining the
  boundary of a void. The exact study of the properties of void volumes
  will be studied in a forthcoming paper (Lavaux \& Wandelt 2009, in
  preparation). We use a variant of the Watershed transform \citep{wvf} to define
  the Lagrangian volume of a void. In Lagrangian coordinates, the
  voids occupy half of the volume. So, instead of enforcing strictly
  that we should have complete segmentation of the volume in terms of
  void patches, we impose that voids must correspond only
  to the places that are sources of displacement and not to sink like
  clusters. Contrarily to a pure Watershed algorithm, we thus enforce that $\sourcePsi > 0$ 
  everywhere within the void boundary.

\subsection{Ellipticity of a void}
\label{sec:ellipticity}

After having defined the position and the dynamical properties of the void, 
we may define an interesting property of those structures: the ellipticity.
\cite{Icke84} first emphasized that isolated voids should evolve to a spherical geometry. But, in the real case, voids are subject to tidal effects. Assuming the present matter distribution evolved from a totally isotropic and homogeneous distribution, 
\cite{ParkLee06} and \cite{LeePark09} have shown that the
distribution of the ellipticity which is produced by tidal effects is a promising probe for cosmology. More generally, previous work have shown that a lot of potentially observable statistical properties of voids are directly to the primordial tidal field \citep[e.g.][]{LeePark06,Platen08,ParkLee09_1,ParkLee09_2}. However,
questions may be raised by the direct use of the formula \cite{Dor70}, 
as they applied it in Millennium Simulation. Using the orbit reconstruction
procedure, our approach should be able
to treat the problem right from the beginning, even in redshift space
(see \citeauthor{lavaux08} \citeyear{lavaux08} for a
long discussion), though some care should be taken for the
distortions along the line-of-sight.

The other advantage of Lagrangian orbit reconstruction is that it
offers for free a way of evaluating the ellipticity {\it locally},
potentially at any space point. From the mass conservation
equation and the definition of eigenvalues of $J_{l,m}$ we may write
the local Eulerian mass density as:
\begin{equation}
  \rho_E({\bf q}) = \frac{\bar{\rho}}{\left|(1 + \lambda_1({\bf q}))(1 + \lambda_2({\bf q}))(1+\lambda_3({\bf q}))\right|} = \frac{\bar{\rho} V_\text{L}}{V_\text{E}({\bf q})}\,,
\end{equation}
with $V_L$ the Lagrangian volume of the cell at ${\bf q}$ and
$V_\text{E}({\bf q})$ the Eulerian volume of this same cell, $\bar{\rho}$
the homogeneous Lagrangian mass density.  This equation is
valid at all times. Now we may also explicitly write the change of
volume of some infinitely small patch of some universe
\begin{equation}
  V_\text{E}({\bf q}) = V_\text{L} \left|(1 + \lambda_1({\bf q}))(1 + \lambda_2({\bf q}))(1+\lambda_3({\bf q}))\right|\,.
\end{equation}
Provided the eigenvalues $\lambda_i$ are greater than -1, which is
always the case for voids, we may drop the absolute value
function.\footnote{An eigenvalue less than -1 would mean that the void
  would have suffered shell crossing at the position of its centre,
  which is dynamically impossible as we are at the farthest distance
  possible of any high density structure.}  Now, with the analogy of
the volume of an ellipsoid,\footnote{We used here the convention of \cite{ParkLee06} who take the square root of the ratio to define $\mu$ and $\nu$.} we may write the ratio $\nu$ between the
minor axis and the major axis
\begin{equation}
  \nu({\bf q}) = \sqrt{\frac{1 + \lambda_3({\bf q})}{1 + \lambda_1({\bf q})}}\,, \label{eq:nu}
\end{equation}
 and the ratio $\mu$ between the second major axis and the
major axis
\begin{equation}
  \mu({\bf q}) = \sqrt{\frac{1 + \lambda_2({\bf q})}{1 + \lambda_1({\bf q})}}\,. \label{eq:mu}
\end{equation}
This allows us to define the ellipticity 
\begin{equation}
 \varepsilon({\bf q}) = 1 - \nu({\bf q}) = 1 - \sqrt{\frac{1 + \lambda_3({\bf q})}{1 + \lambda_1({\bf q})}}\,. \label{eq:epsilon}
\end{equation}
We will define the ellipticity of a void as the value taken by
$\varepsilon$ at the Lagrangian position of the void.

A picture of the concept of voids and ellipticities in this work is
given by Fig.~\ref{fig:transformations}. The painted paraboloid
represents a small piece of a larger 2D density field whose value is
encoded in the height and the colour. According to our definition, the
void is at the centre of the paraboloid. At this centre, the surface of the volume element is mostly circular. The tidal
forces are locally transforming the shape of this surface, which produces
the new elliptic shape on the right side of the figure. The surface has been
here extended along one direction and slightly compressed along the
other. 

Though we are not
strictly limited to study ellipticity only at the position of the void
it may be promising in terms of robustness to non-linear
effects. Indeed, due to the absence of shell-crossings inside voids, the
MAK reconstruction should give the exact solution
\citep{Brenier2002,moh2005,lavaux08} to the orbit reconstruction
problem. It means that the ellipticity that we will compute will be
exact, to the extent that we have taken care of the other potential
systematics due to observational effects \citep{lavaux08}. As any
other method relying on dark matter distribution, we will be sensitive
to the fact that the large-scale galaxy distribution is potentially
biased. However, if the bias does not depends wildly on redshift, we
should be able to compute statistics on ellipticities and derive the
evolution of the growth factor of Large Scale
Structures. 

We note that, using MAK reconstruction, we have access to the
joint distribution of the three eigenvalues.  Our computation of the
ellipticity consists in a projection of the whole joint 3d joint
distribution on a 1d variable. For cosmological analysis, it is not
entirely clear which estimator is the more robust. On one hand, our
intrinsic variables are the eigenvalues and we could include them in the
analysis just as well as the ellipticities
On the other hand, using ellipticity may be helpful to average on a
lot of different voids. It may be a more robust estimator with respect to badly
modelled tails of the distribution of eigenvalues. In this work, we
focus on the use of the ellipticity, as defined in Eq.\eqref{eq:epsilon}. 

\subsection{Eulerian ellipticity}
\label{sec:euler_epsilon}

We define the volume ellipticity $\varepsilon_\text{vol}$ using the eigenvalues of the inertial mass tensor \citep{Shandarin06}:
\begin{equation}
  M_{xx} = \sum_{i=1}^{N_p} m_i (y_i^2 + z_i^2),\hspace{.3cm} M_{xy} =
  - \sum_{i=1}^{N_p} m_i x_i y_i,
\end{equation}
where $m_i$ and $x_i$, $y_i$, $z_i$ are the mass and the coordinates
of the $i$-th particle of the void with respect to its centre of
mass. The other matrix elements are obtained by cyclic permutation of
$x$,$y$ and $z$ symbols.
We put $I_{j}$ the $j$-th eigenvalues of the tensor $M$, with $I_{1}
\leq I_{2} \leq I_{3}$. We may now define the volume ellipticity as in
\begin{equation}
  \varepsilon_\text{vol} =
  1-\left(\frac{I_2+I_3-I_1}{I_2+I_1-I_3}\right)^{1/4} \label{eq:epsilon_shape}
\end{equation}
Even though our work is focused on the tidal ellipticity (Section~\ref{sec:ana_epsilon}), there is some interest to compare how the Eulerian volume ellipticity
compares to the local tidal ellipticity, as most of the existing void finders use $\varepsilon_\text{vol}$
as a probe for the dynamics.

To have a fair comparison with DIVA results, we are computing the inertial
mass tensor from the displacement field ${\bf \Psi}({\bf q})$ smoothed
on the scale as for the rest of the analysis. The void domain is
defined as specified in Sec.~\ref{sec:definition}. The inertial mass
tensor is thus:
\begin{eqnarray}
  M_{xx} & = & \int \text{d}^3 {\bf q} \left((q_{y}+\Psi_{y}({\bf q}))^2 + (q_{z} + \Psi_{z}({\bf q}))^2\right),\\
  M_{xy} & = & - \int\text{d}^3{\bf q} (q_x + \Psi_x({\bf q}))(q_y + \Psi_y({\bf q})).
\end{eqnarray}
with the other elements obtained by cyclic permutations. The volume ellipticity $\varepsilon_\text{vol}$ is compared to the tidal ellipticity $\varepsilon_\text{DIVA}$ in Section~\ref{sec:shapes}. Except in that section, we only consider $\varepsilon_\text{DIVA}$ in this paper.

\subsection{Smoothing scales}
\label{sec:smoothing}

There is an apparent price to pay to go to Lagrangian coordinates. One
has to set a smoothing scale in Lagrangian coordinates and study the
dynamics at corresponding mass scale and let go of the evident notion
of whether we see a hole in the distribution of galaxies or not. 
It actually could be an advantage. Smoothing at different
Lagrangian scales allows to probe the structures at different
dynamical epoch of the void formation. Each Lagrangian smoothing scale
corresponds to a different collapse time: the smallest scales being the fastest
to evolve. DIVA in this respect allows us to study the dynamical properties of
a the voids which have the same collapse time. 

This approach is related to 
the peak patch picture of structure formation \citep{Bond96}, which 
 is a simplified but quite accurate model of the
dynamic of peaks in the density field. This model is even more
precise for the void patches, which is the name of the equivalent model
for studying voids \citep[see e.g.][]{SahniShandarin94,ShethWeygaert04,Novikov06}. 
Of course, the number of voids depends on the filtering
scale (see Section~\ref{sec:analytics_cosmology} and
Section~\ref{sec:mak_vs_sim}). If we smooth on large scales we should
erase the smaller voids and leave only the voids whose size is large
enough. 

 Smoothing also affects the ellipticity distribution. As we
smooth to larger and larger scales the density distribution probed by
the filter should become more and more isotropic. This leads voids to
become more spherical and thus the ellipticity distribution should be
pushed towards a perfect sphere. In
this paper, we consider a few scales separately and try to
understand what were the properties of the minima at each of these
scales (see Section~\ref{sec:nbody_sample}).



\section{Analytical models for voids} 
\label{sec:analytic_voids}

In this section, we describe an analytical model of the displacement
field.  This model is based on Zel'dovich approximation
\citep{zeldov70}. In a first step
(Section~\ref{sec:displacement_stat}), we recall the statistics of the
shear of the displacement field. Then, in
Section~\ref{sec:ana_epsilon}, we express the ellipticity defined by
Eq.~\eqref{eq:epsilon} in terms of this statistic. Finally, we
explicitly write the required statistical quantity in the model of
Gaussian random fields and give some expected general properties of
the voids in this model in Section~\ref{sec:analytics_cosmology}.

\subsection{Displacement field statistics}
\label{sec:displacement_stat}

\cite{ParkLee06} described an analytical model of void ellipticities based on the Zel'dovich
approximation. This model should be particularly suitable on making
predictions of the result given by DIVA, knowing the previous successes of
MAK in this domain \citep{moh2005,lavaux08}. The model that
\cite{ParkLee06} have proposed is based on the unconditional joint
distribution of the eigenvalues of the tidal field matrix $J_{l,m}$ \citep{Dor70},
given the variance of the density field $\sigma^2$
(Appendix~\ref{app:pdf_lambda}):
\begin{multline}
  P(\lambda_1,\lambda_2,\lambda_3|\sigma)  =\\
   \frac{3375}{8\sqrt{5}\sigma^6 \pi} \text{exp}\left[\frac{3\left(2 K_1^2 - 5 K_2\right)}{2\sigma^2}\right] |(\lambda_1-\lambda_2)(\lambda_1-\lambda_3)(\lambda_2-\lambda_3)|\,.
\end{multline}
with
\begin{eqnarray}
  K_1 & = & \lambda_1 + \lambda_2 + \lambda_3\,, \\
  K_2 & = & \lambda_1 \lambda_3 + \lambda_1 \lambda_2 + \lambda_2 \lambda_3 \,.
\end{eqnarray}
as defined in Appendix \ref{app:pdf_lambda}.

This expression however neglects the fact that voids correspond to
maxima of the source of displacement.\footnote{In terms of primordial
  density fluctuations, voids correspond to minima of the density
  field. As MAK is providing a good approximation of this field, we
  may safely jump from one concept to the other.} As the curvature of
$\sourcePsi= \lambda_1+\lambda_2+\lambda_3$ is correlated with
$J_{l,m}$, we need to enforce that we are actually observing the
eigenvalues in regions where the curvature of $\sourcePsi$ is
negative. A better expression would be derived if we could constrain
that the Hessian $H$ (the matrix of the second derivatives) of $\sourcePsi$ is negative,
which is the case in the vicinity of maxima of $\sourcePsi$, the
source of the displacement field. We derive in
Appendix~\ref{app:void_tidal} a general formalism that allows us to
compute numerically the probability
$P(\lambda_1,\lambda_2,\lambda_3|\sigma_T,r,H<0)$ to observe the
eigenvalues $\{\lambda_1,\lambda_2,\lambda_3\}$ given that we look in
these regions. This formalism is a natural extension of the formula of
\cite{Dor70} (for which a simple derivation is given in
Appendix~\ref{app:pdf_lambda}).

Of course, ``true voids'' have the additional constraint
that $\lambda_i > 0$ for all $i=1,2,3$.  As we assumed in previous sections
that eigenvalues are ordered according to $\lambda_1 > \lambda_2 > \lambda_3$,
the constraint $\lambda_3 > 0$ is sufficient to study this case. 

\subsection{Ellipticity statistics}
\label{sec:ana_epsilon}

Whether we use the conditional probability
$P(\lambda_1,\lambda_2,\lambda_3|\sigma_T,r,H<0)$ or the unconditional
one $P(\lambda_1,\lambda_2,\lambda_3|\sigma_T)$, both under 
notation $\mathcal{P}(\lambda_1,\lambda_2,\lambda_3|\sigma_T,r)$,
we may now express the
probability to observe $\delta$, $\nu$, $\mu$ [defined in
  Equations~\eqref{eq:delta}, \eqref{eq:nu} and \eqref{eq:mu}] in
terms of $\mathcal{P}$:
\begin{equation}
  P(\mu,\nu,\delta|r,\sigma_T) = \mathcal{P}(\lambda_1, \lambda_2, \lambda_3|r,\sigma_T) \times \frac{4 (\delta-3)^2 \mu\nu}{(1+\mu^2+\nu^2)^3}\,, \label{eq:axis_distrib}
\end{equation}
with
\begin{eqnarray}
  \lambda_1 & = & -\frac{1 + \mu^2 - 2 \nu^2 + \delta \nu^2}{1 + \mu^2 + \nu^2}\,, \\
  \lambda_2 & = & -\frac{1 - 2\mu^2 + \delta \mu^2 + \nu^2}{1 + \mu^2 + \nu^2}\,, \\
  \lambda_3 & = & -\frac{-2 + \delta + \mu^2 + \nu^2}{1 + \mu^2 + \nu^2}\,.
\end{eqnarray}
The ellipticity distribution of voids is thus
\begin{equation}
  P(\varepsilon|\sigma_T,r) = \frac{1}{\mathcal{N}}\int_{\delta=-\infty}^{+\infty}\int_{\mu=1-\varepsilon}^1 P(\mu,\nu,\delta|\sigma_T,r)\,\text{d}\mu\text{d}\delta ,
\end{equation}
with
\begin{equation}
  \mathcal{N} = \int_{\delta=-\infty}^{+\infty} \int_{\nu=0}^1\int_{\mu=\nu}^1\text{d}\mu\text{d}\nu\text{d}\delta\,P(\mu,\nu,\delta|\sigma_T,r)\,.
\end{equation}
The alternative distribution for ``true voids'' is given by enforcing
that $\lambda_1 > 0$ and may be obtained by introducing the Heaviside
function $\Theta(\lambda_1)$ in Eq.~\eqref{eq:axis_distrib} and renormalising.

We note that the ellipticity that we are considering here
  is of dynamical nature \citep[as emphasized by][]{ParkLee06}. This
  comes in contrast with the first studies of void ellipticities due
  to redshift distortions by \cite{Ryden95} and
  \cite{MelottRyden96}. We do not discuss this earlier definition of
  ellipticity but only the later one.

\subsection{Application to cosmology}
\label{sec:analytics_cosmology}

{\it Shapes of voids --}
Now we may compute the ellipticity distribution of voids,
$P(\varepsilon|\sourcePsi)$ for a given cosmology.
 The variance of the density field
$\sigma_T^2$, assuming the power spectrum of the density fluctuations
$P(k)$, is given by
\begin{equation}
  \sigma^2_T = \frac{1}{2\pi^2} \int_0^{+\infty} P(k) W^2_{R_f}(k)\,k^2\text{d}k\,,\label{eq:var_j}
\end{equation}
with 
\begin{equation}
  W_{R_f}(k) = \exp\left(-\frac{k^2}{2 R_f^2}\right)
\end{equation}
the Fourier transform of the filter function used to
smooth the density field on the scale $R_f$.
In practice, we smooth the displacement field in Lagrangian
coordinates, to reduce noise and non-linear effects appearing at small
scales in the MAK reconstruction ($\la 5$\Mpch). Thus, we will compute
the ellipticity distribution of voids, given that we smoothed on the
scale $R_f$ in Lagrangian coordinates, and that the local source of 
displacement of the void is $\sourcePsi({\bf q})$.

With the model $P(\lambda_1,\lambda_2,\lambda_3|\sigma_T,r,H<0)$ of Appendix~\ref{app:void_tidal}, we may also
estimate the number of voids in each class we defined in Section~\ref{sec:definition}. As an illustration, we picked a usual \LCDM{} cosmology, with $\Omega_\text{m}=0.30$, $\Omega_\text{b}=0.04$, $\sigma_8=0.77$,
  $h=0.65$ and estimated the fraction of voids in each class. The results are, when we smooth at 4\Mpch{}:
\begin{itemize}
\item {\it  true voids}:
  We estimate that these voids represent
  $\sim$40\% of the primordial voids, which correspond to underdensities
  in the primordial density fluctuations.
\item {\it pancake voids}:
  Doing the same estimation as for ``true voids'', we get that
  $\sim$50\% of the primordial voids should be in that case.
\item {\it filament voids}: They correspond to $\sim$10\% of the primordial
  voids.
\end{itemize}

\begin{figure*}
  \includegraphics[width=\hsize]{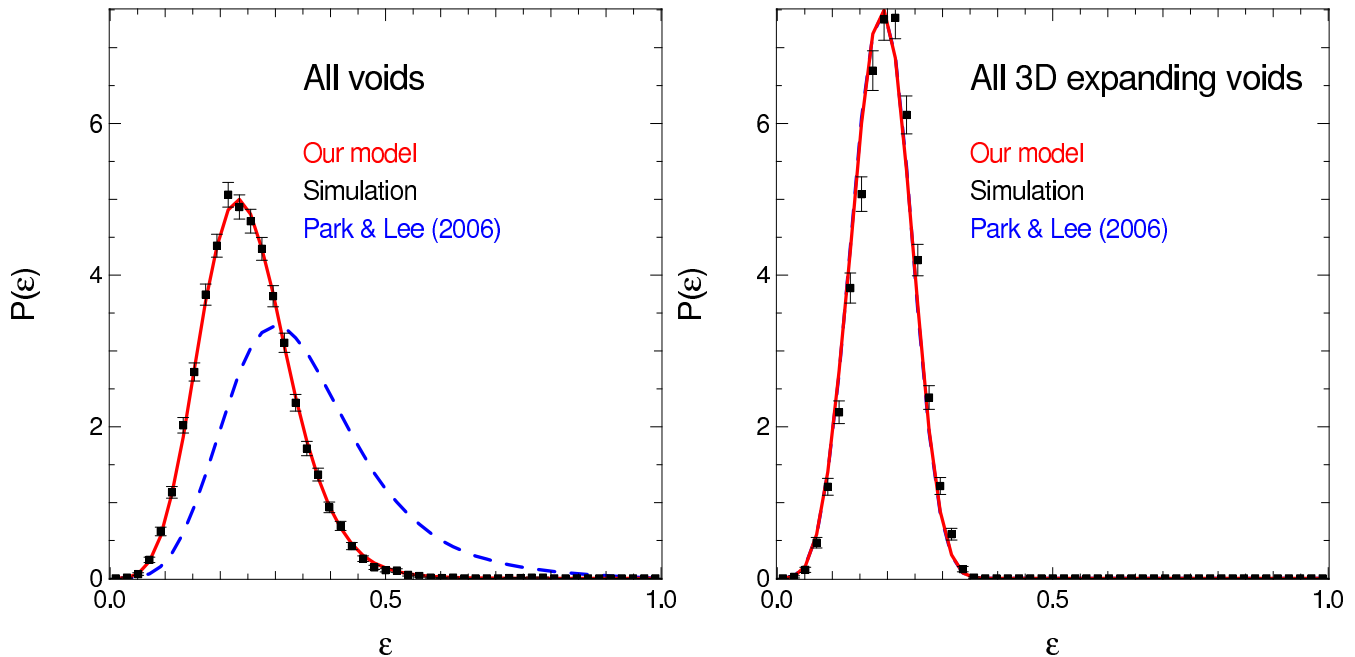}
  \caption{\label{fig:ellipticity} {\it Comparison of MAK reconstructed and
      analytical ellipticity distribution --} We represent here the
    distribution of ellipticity of the voids, marginalised over all
    possible $\sourcePsi$. We used black squares for the
    ellipticity distribution obtained using the MAK reconstructed
    displacement field applied
    on the simulation. The dashed blue curve is computed using the
    unconditional \protect\cite{Dor70} formula. The red curve is our new
    formula obtained by conditioning that voids are regions where the
    density field has a positive curvature. The left panel gives the
    result for all voids (true, pancake and filament). The right panel
    gives the result for only true voids. The blue dashed curve and red solid
    curve overlap. All fields were smoothed with a Gaussian kernel of radius 4\Mpch{}. }
\end{figure*}

We show in Figure~\ref{fig:ellipticity}, the analytical
distributions of ellipticity for the two cases when one constrains (or not) the curvature
of $\sourcePsi$ to be negative. The solid curve corresponds to the
ellipticity distribution obtained using
$P(\lambda_1,\lambda_2,\lambda_3|H<0)$. The dashed curve is obtained
with the unconstrained distribution. In the left panel, we took all
voids with $\lambda_1 > -1$. In the right panel, we restricted
ourselves to ``true voids''. The difference is striking in the left
panel between the two models, whereas in the right panel they
essentially give the same prediction. This can be understood by looking
at the value of the correlation coefficient between the curvature of
$\sourcePsi$ and $T_{l,m}$ (also defined in
Eq.~\eqref{eq:correl_curvature})
\begin{equation}
  r = -\frac{\mathcal{S}_4}{\sqrt{\mathcal{S}_2 \mathcal{S}_6}}\,,
\end{equation}
with
\begin{equation}
  \mathcal{S}_n = \frac{1}{2\pi^2} \int_{k=0}^{+\infty} k^n
  P(k)\,\text{d}k \label{eq:var_s_n}
\end{equation}
This coefficient is equal to $\sim$-0.67 for the aforementioned
cosmology. This indicates that the two curvatures are quite strongly
linked. Thus, enforcing that $T_{l,m}$ is positive causes 
the curvature of $\sourcePsi$ to be preferentially 
negative.  So, the two distributions of the right panel of
Fig.~\ref{fig:ellipticity} should be similar. On the
other hand, for the left panel no such implication exists, which leads
to the largely evident discrepancies of ellipticities. {\it We note the
distributions of the right panel are only measurable using our void
finder as the other ones cannot distinguish truly expanding voids and
pancake voids just by looking at their shape.}

{\it Number of voids --}
Now that we know the shape the voids should have in the context of linear theory, we would like to know how many of them should be present in the Universe. 
With our definition of voids, we may conveniently use the results obtained by \cite{BBKS} for Gaussian random fields. In particular in the equation~(4.11b), they show that the average number density of maxima is equal to
\begin{equation} 
  n_\text{max} = \frac{29-6\sqrt{6}}{5^{3/2} 2 (2\pi)^2 R_*^3} \simeq 0.016 R_*^{-3}\,. \label{eq:num_maxima}
\end{equation}
with
\begin{equation}
  R_* = \sqrt{3 \frac{\mathcal{S}_4}{\mathcal{S}_6}}
\end{equation}
in our notation.
We note that this is a mean number, so we expect some fluctuation according to
the mean which are  difficult to compute analytically. Tests on Gaussian random field seems to point out that the number of voids are distributed like a Poisson distribution. We
also expect this number to slightly overestimate the actual density
of voids that we will find in simulations (Section~\ref{sec:mak_displacement}).

In the next section, we are now going to confront the analytical model
with the results given by DIVA applied on $N$-body simulations.


\section{Test on $N$-body simulations}
\label{sec:nbody_test}

In this section, we are going to identify voids in the $N$-body
samples described in Section~\ref{sec:nbody_sample}. We then give a
sketch (Section~\ref{sec:mak_displacement}) of the procedure we
followed to perform the MAK reconstruction, which corresponds to the one
described in \cite{lavaux08}. In Section~\ref{sec:results_z0}, we
focus on the results obtained at $z=0$.  First, we give an illustration of a void of each class in Section~\ref{sec:examples}. We then compare 
the results obtained using the displacement field given by the
simulation and the one reconstructed by MAK
(Section~\ref{sec:mak_vs_sim}). There, we also detail the number of
voids detected and their ellipticities for both fields. We compare 
quantitatively the distribution of Eulerian volume ellipticity to
the Lagrangian tidal ellipticity in Section~\ref{sec:shapes}. Finally, we
check the validity of the analytical model on the simulated
displacement field (Section~\ref{sec:simu_vs_analytic}). In
Section~\ref{sec:evolution}, we look at the evolution of the mean
ellipticity for a simulation with $w=-1$ and in the analytical
model. At last, in Section~\ref{sec:two_cosmologies}, we assess the
possibility of measuring the evolution of the mean ellipticity in two
simulations where $w$ is either $-1$ or $-0.5$.

\subsection{The $N$-body simulations}
\label{sec:nbody_sample}

To test our void finder, we use three large volume $N$-body simulations but
with medium resolution of $N=512^3, L=500$\Mpch,
$\Omega_\text{m}=0.30$, $\Omega_\Lambda=0.70$, $H=65$~\kmsMpc,
$n_s=1$, $\sigma_8 = 0.77$, $\Omega_\text{b}=0.04$. The $N$-body
simulations contains only dark matter and we include the effect of baryons
only through power spectrum features in the initial conditions.  This
essentially reduces power on scales smaller than the sound horizon and
introduces Baryonic Acoustic Oscillations. The two first $N$-body simulation
($\Lambda$SIM and $\Lambda$SIM2) corresponds to a standard \LCDM{} cosmology
for which the equation of state is given by $w=-1$. $\Lambda$SIM and
$\Lambda$SIM2 have exactly the same cosmology but have different
initial conditions. They will allow us to assess the impact of
looking at two different realisations of the initial conditions.
 The third, wSIM, is assuming an equation of state $w=-0.5$ for the Dark
Energy.  To build the initial conditions, we modified the
\verb,GRAFIC, \citep{grafic2} package to use the power spectrum generated by the
\verb,CAMB, package \citep{Lewis:1999bs}. We reach a sub-Mpc resolution scale which
is sufficient for proper description of most voids (1-20 \Mpch). The
intermediately large volume allows accounting for
large-scale tidal field effects and cosmic variance effects. We used the parallelised version of
the \verb,RAMSES, $N$-body code \citep{ramses} and run it both on the
Cobalt NCSA supercluster and the Teragrid NCSA supercluster through
Teragrid facilities \citep{teragrid}.  We modified \verb,RAMSES, to
simulate cosmologies with a dark energy equation of state different
than $w=-1$. 

To account for the impact of clustering, we analyse the
displacement field for which the mass of dark matter halos has been
put to the centre of mass of these halos. To be able to do that, we
construct a halo catalogue using a Friend-of-Friend algorithm with a
traditional linking value of $l=0.2$ \citep{Davis85,EFWD}. We put a
threshold of $N_\text{p} = 8$. This prescription in practice should
mostly erase the information contained in halos while retaining the
dynamics outside of them. Though the use of such a low number for the
particles of halos are questionable for the study of the properties of 
those halos, we are not here interested in them. We are only interested
in checking that we keep most of the information useful to study voids
and their overall dynamics, even though we destroy the small scale information. The above prescription has already been
successfully applied for the study of peculiar velocities with MAK
\citep{lavaux08}. We note that we will only use the halo catalogue to do the
MAK reconstruction. All the tests of the displacement field of the simulation
are run on the {\it particles} of the simulation. 

We note that that the 
initial conditions of the simulation present two particularities that must be
taken into account. The largest mode of the simulation box is
$k_\text{low}=1.25\times 10^{-2}\,h$~Mpc$^{-1}$ thus introducing a sharp
low-k cut.  Additionally \verb,GRAFIC, applies a Hanning filter on the
initial conditions to avoid aliasing. In practice, \verb,GRAFIC, multiplies
the cosmological power spectrum by the following filter:
\begin{equation}
  W_\text{hanning}(k) = \left\{
  \begin{array}{ll}
    \left[\cos\left(\frac{1}{2} k \Delta x\right)\right] & \text{if } k \Delta x \le \pi \\
    0 & \text{elsewhere}
  \end{array}
  \right.,
\end{equation}
with $\Delta x = 0.976$\Mpch{} the Lagrangian grid step size of our simulations.
  These two features must be
introduced in the power spectrum of Eq.~\eqref{eq:var_j} and
\eqref{eq:var_s_n} to make correct predictions for the simulation. In
real observational data, no such features should be present.

\subsection{MAK reconstruction and void identification}
\label{sec:mak_displacement}

Among the different tests that we are going to present in the
following, we have run only one MAK reconstruction for the full halo
catalogue. We chose a resolution of $256^3$ mesh elements generated
following the algorithm of \cite{lavaux08}. This algorithm consists in
splitting a mass in an integer number of MAK mesh elements, with the constraints
that the splitting must be fair and the number of mesh elements is fixed
and equal to $256^3$. This method works well and has, so far, not 
been prone to biases.
Choosing this number of elements gives us a resolution
of $\sim$2\Mpch{} on the Lagrangian coordinates of the displacement field.
We cannot run it at the full simulation resolution due to the high
CPU-time cost which hinders doing several reconstruction.
One reconstruction takes  $\sim$13,000
  accumulated CPU-hours on Teragrid cluster at NCSA. However, as the
  complexity grows as $O(N^{2.25})$, increasing the resolution to $512^3$
  would have consumed $\sim 10^6$ CPU-hours. So we limited ourself on
  running the reconstruction at the $256^3$ resolution, the expected
  worst case for the performance of MAK. At higher redshift, the MAK
  reconstruction converges faster and gives a more and more precise
  displacement field compared to the one given by simulation. These
  two features are both caused by the decrease of small scale
  non-linearities at earlier times. We took the halo catalogue built
  on $\Lambda$SIM at $z=0$
and ran a reconstruction on it. The other results presented in this
paper use the displacement field given directly by the simulation
after having checked that the reconstruction is indeed sufficient for
our purpose. This is the case as we will not look at voids smaller
than $\sim$1\Mpch{} scale in Lagrangian coordinates. 

We chose two different smoothing scales on which we study the
displacement field for voids: 2.5\Mpch{} and 4\Mpch{}. Once the
displacement field has been smoothed, we compute the eigenvalues and
the divergences in Fourier space. We then locate the maxima in the
divergence of the displacement field. At each maxima, we compute the
ellipticity $\varepsilon$ with the help of Eq.~\eqref{eq:epsilon}, where
the $\lambda_i$ are taken as the eigenvalues of $T_{l,m}({\bf q})$. The 
displacement shear tensor, is computed numerically from ${\bf \Psi}({\bf q})$ in Fourier space:
\begin{equation}
  T_{l,m}({\bf q}) = \frac{1}{(2\pi)^3} \int_{\bf k} \text{d}^3{\bf k} i k_m \hat{\Psi}_l({\bf k}) \text{e}^{i {\bf k}\cdot {\bf q}},\label{eq:fourier_tidal}
\end{equation}
where $\hat{\Psi}_l({\bf k})$ is the Fourier transform of the displacement 
field in Lagrangian coordinates. All these quantities were evaluated using
Fast Fourier Transform on the Lagrangian grid.

In summary, the steps are the following:
\begin{itemize}
\item[-] First, we prepare a catalogue for a MAK reconstruction. This involves doing fair equal mass-splitting of the objects.
\item[-] Next, we run the MAK reconstruction.
\item[-] After the reconstruction is finished, we put the computed displacement field given by MAK on the homogeneous Lagrangian grid.
\item[-] We smooth this displacement field and compute the tidal field $T_{i,j}$ in Lagrangian coordinates in Fourier space using Eq.\eqref{eq:fourier_tidal}. 
\item[-] Now we compute $\sourcePsi({\bf q})$ on the grid using Eq.~\eqref{eq:delta}, which corresponds to summing the three eigenvalues of $T_{i,j}$.
\item[-] We look for local maxima in $\sourcePsi({\bf q})$ using an iterative steepest descent algorithm. This gives us the list of the voids in Lagrangian coordinates.
\item[-] Using these coordinates, we now compute the ellipticity using the eigenvalues of $T_{i,j}$ at the location of the minima.
\item[-] For computing the void boundary, we execute the modified Watershed transform of Section~\ref{sec:definition}. This identifies the Lagrangian domain of the voids. The boundary of this domain is then transported 
using the displacement field to recover the Eulerian void volume.
\end{itemize}
We now look at the results
obtained with MAK, the simulation and the analytical model at $z=0$ in the next
section.

\subsection{Results at $z=0$}
\label{sec:results_z0}

\subsubsection{Example of found voids}
\label{sec:examples}

\begin{figure*}
  \includegraphics[width=\hsize]{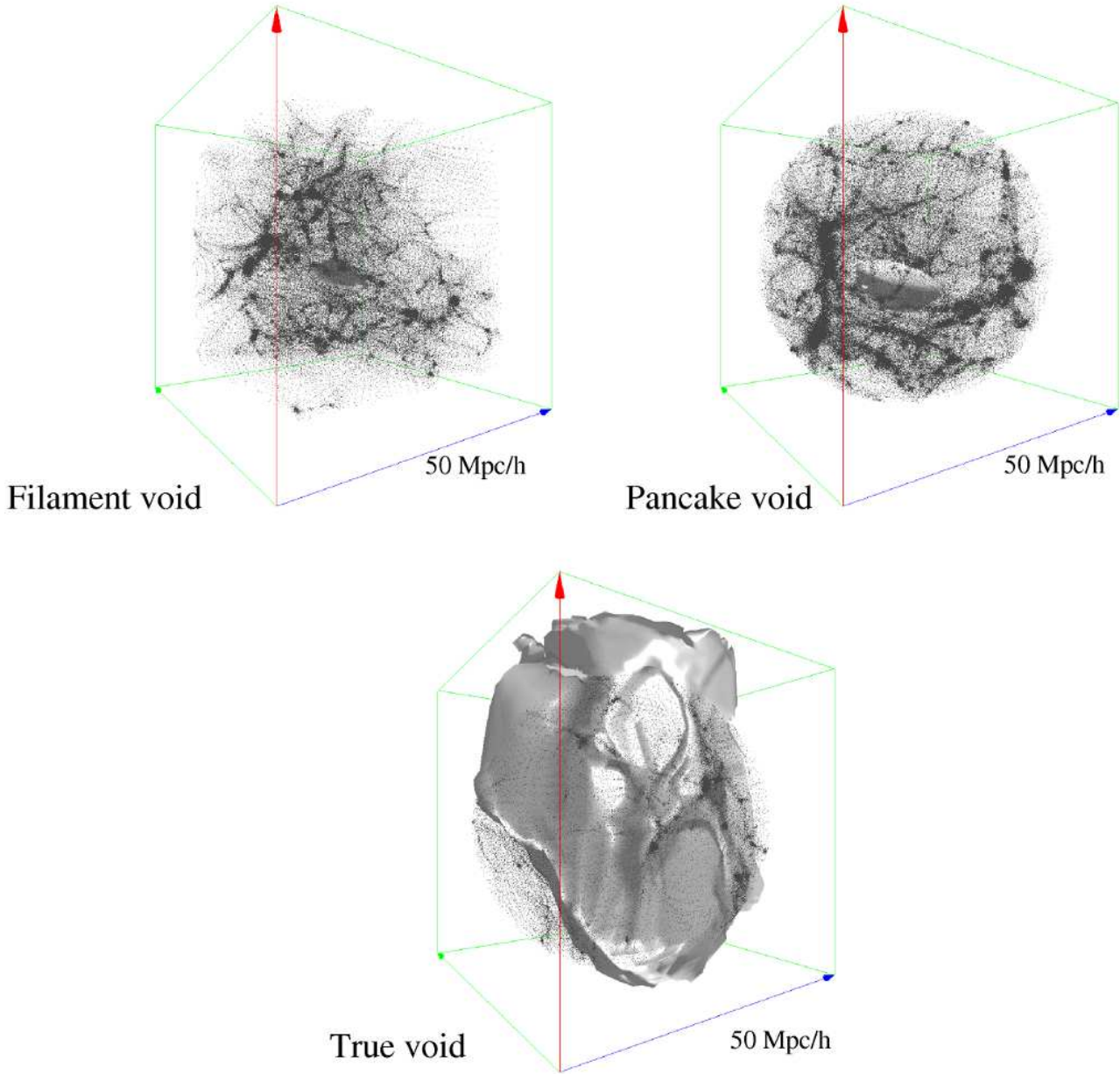}
  \caption{\label{fig:divavoid} {\it Example of voids} -- We illustrate
  here the voids that are found using DIVA. Each of these belong to one of
  the void class that we listed in Section~\ref{sec:definition}. The scale of the box is the same for the three cases: the side corresponds to 50\Mpch{}. }
\end{figure*}

Before going to the statistical study of the local shape $\varepsilon_\text{DIVA}$
  properties of void found by DIVA, we look at a visual example of
  each of the void type. We chose a filtering scale of 4\Mpch{} in
  Lagrangian coordinates. We selected one
  void of each class. These three voids have the following
  properties:
\begin{itemize}
\item The first void is a true void. The eigenvalues
  of the tidal field $T_{l,m}({\bf q})$ \eqref{eq:tidal_def} at the
  centre are $(1.2,0.84,0.69)$ along the three axis. We thus derive
  the ellipticity $\varepsilon = 0.13$. The volume of the void, in
  Lagrangian coordinates, is $V_\text{L} = 75240$\VolMpch, which
  corresponds to an equivalent spherical volume given by a sphere of
  radius $R_\text{equiv} = (3/(4\pi)*V)^{1/3} = 26$\Mpch{}.
\item The second void is a pancake void. The
  eigenvalues of the tidal field are $(1.1,0.11,-0.60)$, the
  ellipticity is $0.563$ and the Lagrangian volume is $V_\text{L}= 1560
  h^{-3}$~Mpc$^{3}$, with $R_\text{equiv} = 7.2$\Mpch{}.
\item The last void is a filament void. Again, the
  eigenvalues of the tidal field are $(0.99,-0.24,-0.61)$, the
  ellipticity is $\varepsilon=0.557$ and the Lagrangian volume is
  $V_\text{L}= 260 h^{-3}$~Mpc$^{3}$, with $R_\text{equiv} =
  4.0$\Mpch{}.
\end{itemize}
Those three voids are represented in three dimensions, along with the
particles of the simulation in the same region, in
Fig.~\ref{fig:divavoid}. We note that the true voids is the largest one. We expect to observe that
effect as true voids expands in three directions and thus are more likely to be bigger than pancake voids and filament voids. For these three cases, the surface delimited by DIVA seems to nicely fit to the structures located on the boundaries. In the case of the pancake and filament voids, we note that the cavity seem to be split into two pieces. This splitting is due to the Watershed transform criterion which isolated two void cavities separated by a structure. 

\subsubsection{MAK vs Simulation}
\label{sec:mak_vs_sim}

\begin{figure*}
  \includegraphics[width=\hsize]{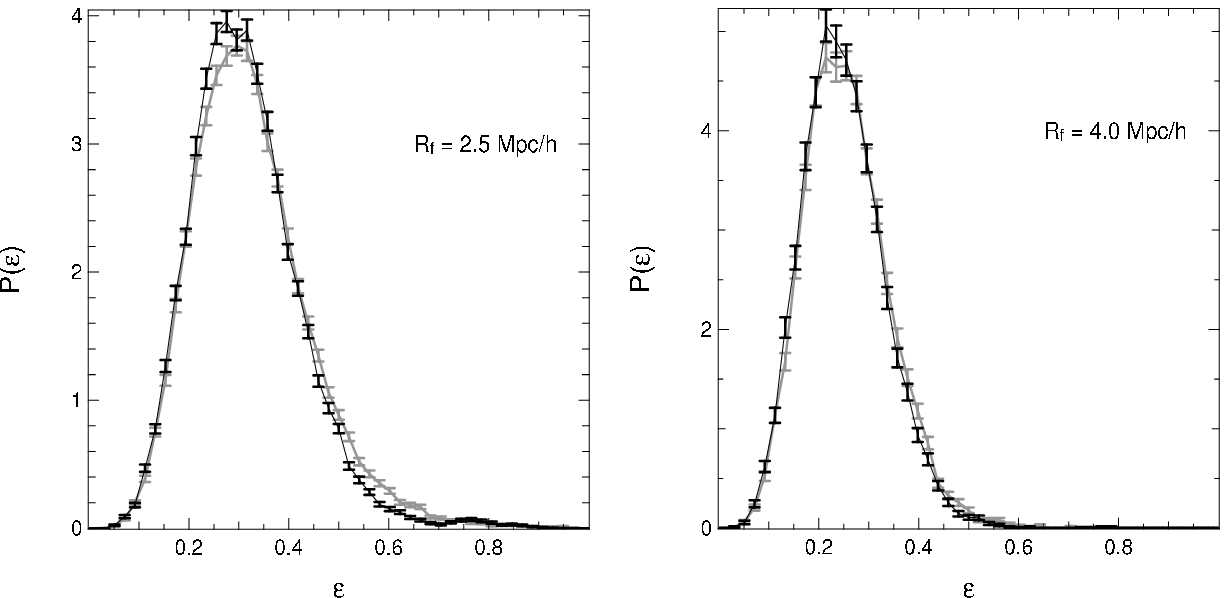}
  \caption{\label{fig:comparison_simu_vs_mak} {\it MAK reconstructed ellipticity
      distribution vs. ellipticity distribution from simulated displacement field} --
    This figure gives the
    ellipticity distribution computed using either the MAK
    reconstructed displacement field (solid black line) or the
    simulated displacement field (solid thick gray line). We
    represent the dispersion expected if the error on estimating the
    distribution come from the number of voids in a given bin. We
    assumed a Poisson distribution for the estimation of the error bar. The
    displacement fields were both smoothed with a Gaussian kernel of
    radius $R_f=2.5$\Mpch{} in the left panel and to $R_f=4$\Mpch{} in the right panel.}
\end{figure*}

We now concentrate on the properties of the voids at $z=0$.  This is
the time where the density is the most non-linear and the
reconstruction is the most difficult and therefore represents a worst
case example. We take the displacement field given
by the simulation as the field of reference to study voids. Indeed,
this field has been obtained by solving completely the equation of
motion for each particle. In this section, we will first compare the
properties of the voids found using this field and the MAK
reconstructed field. Then, we will focus on how it compares to the
analytic model.

We represent on Fig.~\ref{fig:comparison_simu_vs_mak} the
distribution of ellipticities obtained using both the reconstructed
and the simulated displacement field.  We also give the number of
voids found in simulated displacement field, in the reconstructed
displacement field and the expected number of maxima according to
Eq.~\eqref{eq:num_maxima} (Table~\ref{tab:num_voids}). 
To allow for a void-by-void comparison, we tried to match the voids found
using the two displacement fields. We considered that two voids are the
same if the distance between the two voids is less than a Lagrangian grid cell ($d \le \sqrt{3}\,l_\text{cell}$, with $l_\text{cell}$ the length of the side of a cell). 

At 2.5\Mpch{} smoothing, the agreement of the ellipticity distribution
derived from the simulated displacement and the MAK reconstructed
displacement is very good, though the high ellipticity tail seems a
little different in the two cases. This is actually due to a selection
effect which, unfortunately, is correlated with the ellipticity. If we
look at the number of voids detected using the two fields
(Table~\ref{tab:num_voids}), we see that MAK is missing about 10\% of
the voids detected in the simulation.  The distribution of ellipticity
of those voids happens to be skewed towards higher ellipticities. Thus
it seems that we tend to miss the most elliptical voids.  This
behaviour is expected as these voids tend to be pancake voids. So
they are closing along one direction, and the more elliptical they are the
faster they are closing. If they close, the particles of these voids
 shell cross and MAK is not able to reconstruct the
displacement field.  Looking at this same distribution, but for a
4\Mpch{} smoothing, we now barely see a difference between the two
curves. We indeed checked that the ellipticity distribution of the
voids that have not been identified using the MAK reconstructed
displacement field is similar to the distribution of the
other voids. 

The number of voids detected
in the simulation looks less than the expected average number of minima
(Table~\ref{tab:num_voids}). This is also due to the destruction of minima
by the collapse of large scale structures. When we look at the beginning of
the simulation we find 11,485 minima (for $R_f=4$\Mpch), and this number steadily decreases to 
the value we put in the table as the simulation evolves. We estimated
the mean and the variance in the number of minima using 
20 realisations of a Gaussian random field with the same cosmology
as the simulation.
We found that the mean should be 11,762 and the standard deviation is 58, which is in agreement with the result given by the analytic computation.

\begin{table}
  \begin{center}
    \begin{tabular}{cccc}
      \hline
      \hline
      \multirow{2}{1cm}{Filter} & Predicted & Displacement & Number of \\
      & average number & field & candidates \\
      \hline
      \hline
      2.5\Mpch{} & 42908 & Simulated & 31002 \\
      &   & Reconstructed & 28397 \\
      4\Mpch{} & 11706 & Simulated & 10643 \\
      &  & Reconstructed & 9412\\
      \hline
    \end{tabular}
  \end{center}
  \caption{\label{tab:num_voids} {\it Unconditioned number of voids in
      $\Lambda$SIM} -- we give here the unconditioned number of voids
    found within the volume of the simulation, (500 \Mpch)$^3$. The
    predictions are obtained using Equation~\eqref{eq:num_maxima}
    applied on the power spectrum of primordial density fluctuations
    multiplied by the Fourier transform of the indicated filter in the
    first column. The last column gives the actual number of void
    candidates that we found using the displacement field. }
\end{table}

\begin{figure*}
  \includegraphics[width=\hsize]{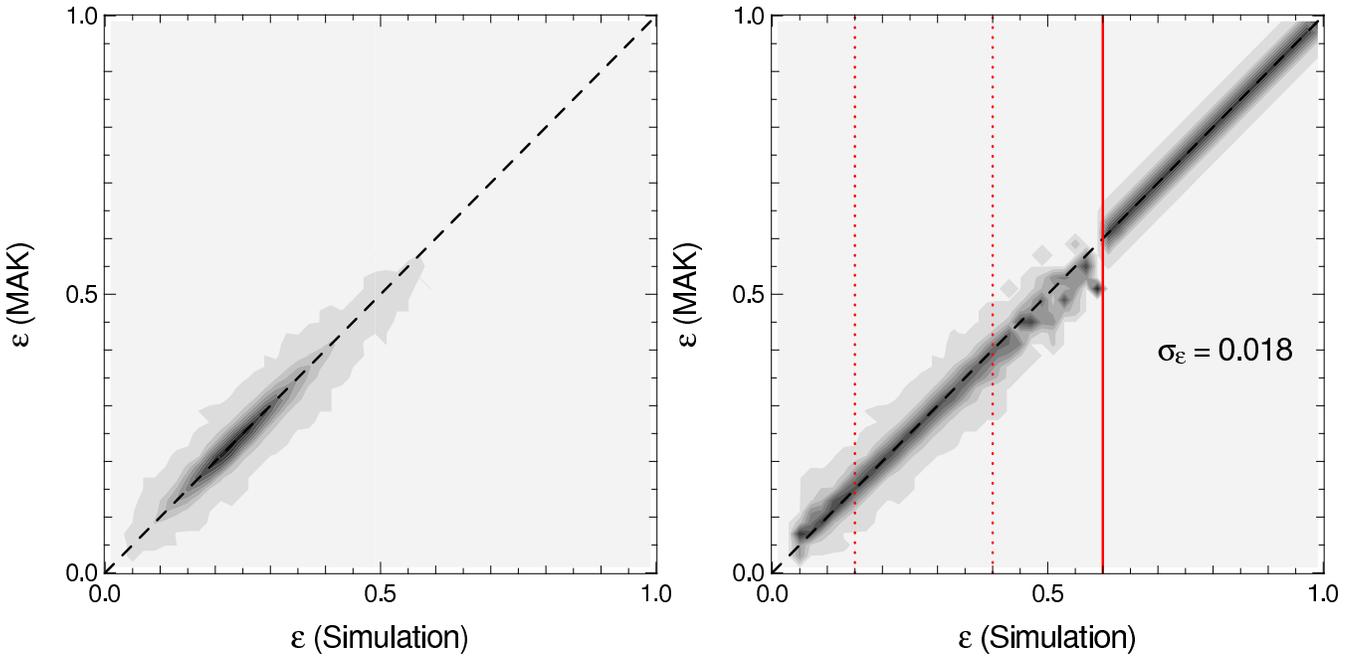}
  \caption{\label{fig:ellipticity_scatter} {\it Ellipticity derived
      from the simulated displacement field vs. the MAK displacement
      field} -- We represent here a scatter plot between the
    ellipticities of the voids that were both detected in the MAK
    reconstructed displacement field and the simulated displacement
    field, both smoothed at 4\Mpch{}. In the left panel, we
    show the raw joint probability distribution of the two
    ellipticities. The gray-scale is linear in the density of probability.
    In the right panel, we represent the conditional
    distribution of $\varepsilon_\text{MAK}$ given
    $\varepsilon_\text{SIM}$, on the left of the thick vertical
    line. On the right of this same line, we represent this same
    distribution if one uses the estimated standard deviation $\sigma_\varepsilon$
    of the error. It is estimated using the distribution between the
    two vertical dotted lines. The estimated standard deviation is $\sigma_\varepsilon = 0.018$.}
\end{figure*}

Using the match between voids coming from the two fields, we can build 
a statistical error model in the form of the joint probability distribution
$P(\varepsilon_\text{MAK},\varepsilon_\text{SIM})$ of getting an
ellipticity $\varepsilon_\text{MAK}$ using MAK displacement and an
ellipticity $\varepsilon_\text{SIM})$ using simulated displacement. It
is important to check
$P(\varepsilon_\text{MAK},\varepsilon_\text{SIM})$ if, as we will do
in future, we want to estimate cosmological parameters from
ellipticity distribution. This function tells us what accuracy we may
expect on the ellipticity measurements.  We represent this
probability in the left panel of
Figure~\ref{fig:ellipticity_scatter}. We see a strong correlation,
already seen in Fig.~\ref{fig:comparison_simu_vs_mak}, indicating a high
accurate reconstruction. We also see
that the error seems low. We represent at the left side of the thick
red solid line of the right panel the conditional probability
$P(\varepsilon_\text{MAK}|\varepsilon_\text{SIM})$ that we computed
using:
\begin{equation}
  P(\varepsilon_\text{MAK}|\varepsilon_\text{SIM}) = \frac{P(\varepsilon_\text{MAK},\varepsilon_\text{SIM})}{\int_{\varepsilon_\text{MAK}=0}^1 P(\varepsilon,\varepsilon_\text{SIM}) \text{d}\varepsilon}
\end{equation}
wherever it was possible to evaluate the denominator. This conditional
probability is mostly Gaussian, so we tried to estimate the standard
error by computing the mean variance of the error on the ellipticity
for the distribution between the two dotted red line
$\varepsilon_\text{SIM} \in [\varepsilon_{S,min};\varepsilon_{S,max}]$
with $\varepsilon_{S,min}=0.15$ and $\varepsilon_{S,max}=0.40$. With
this notation, we computed
\begin{multline}
  \sigma_\varepsilon = \frac{1}{\varepsilon_{S,max}-\varepsilon_{S,min}} \\
  \int_{\varepsilon=\varepsilon_{S,min}}^{\varepsilon_{S,max}}\text{d}\varepsilon \sqrt{ \int_{\varepsilon_\text{\sc mak}=0}^1\text{d}{\varepsilon_\text{\sc mak}} (\varepsilon_\text{\sc mak}-\varepsilon)^2 P(\varepsilon_\text{\sc mak}|\varepsilon)}.
\end{multline}
Within the interval limited by the two dotted red line, we estimate
$\sigma_\varepsilon \simeq 0.018$. At the right of the vertical red solid line,
we show the function
\begin{equation}
  P(\varepsilon|\varepsilon_\text{SIMU},\sigma_\varepsilon) = \frac{1}{\sqrt{2\pi} \sigma_\varepsilon} \text{e}^{-\frac{1}{2\sigma_\varepsilon^2}(\varepsilon-\varepsilon_\text{SIMU})^2}
\end{equation}
with $\sigma_\varepsilon$ as estimated above. We note that this
model of the conditional probability (right of the
vertical solid line) looks much like the actual ellipticity
discrepancy (left of the vertical solid line).

\subsubsection{Volume ellipticity $\varepsilon_\text{vol}$ vs. Tidal ellipticity $\varepsilon_\text{DIVA}$}
\label{sec:shapes}

\begin{figure}
  \includegraphics[width=\hsize]{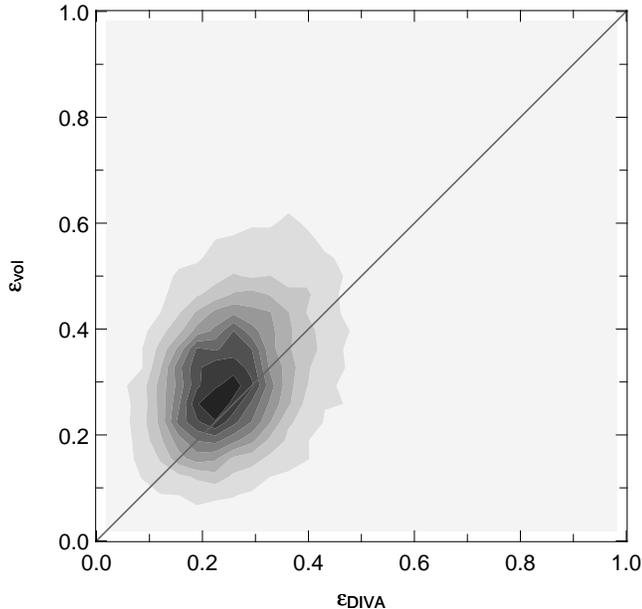}
  \caption{\label{fig:epsilon_comp} {\it Tidal ellipticities vs Volume ellipticity} -- This plot gives a comparison of the ellipticity of the void as determined either using the shear of the displacement field [$\varepsilon_\text{DIVA}$, Eq.~\eqref{eq:epsilon}] or using the overall shape of the void [$\varepsilon_\text{vol}$, Eq.~\eqref{eq:epsilon_shape}].}
\end{figure}

In Fig.~\ref{fig:epsilon_comp}, we represented a
  comparison between the ellipticity of the volume, $\varepsilon_\text{vol}$, and the local tidal
  ellipticity, $\varepsilon_\text{DIVA}$. The volume ellipticity is computed using the
  Eq.~\eqref{eq:epsilon_shape}, with the inertial mass tensor $M$ as
  computed using the smoothed displacement field. Visually, the two
  variables seem loosely correlated. We observe they do follow each
  other but this correlation just get worse when the ellipticity
  increases. This is expected as the volume ellipticity is a non-local
  quantity and thus is sensitive to all local ellipticities in the
  void volume. This is what makes $\varepsilon_\text{vol}$ more difficult to use as a
  precise cosmology probe. We show in
  Appendix~\ref{app:local_vs_global_ellipticity} that these two
  quantities are indeed related but only to first order. This explains that the 
  scatter seems smaller for small ellipticities than for high ellipticities.
  {\it The volume ellipticity, which has been used so far, seems thus to be a poor 
  proxy of the tidal ellipticity, which we manage to recover with extreme
  precision using our Lagrangian orbit reconstruction technique.}  
  For the rest of this paper, we will only use the tidal ellipticity.

\subsubsection{Simulation vs Analytic}
\label{sec:simu_vs_analytic}

In this section, we compare the results obtained on the simulated
displacement field and the prediction given by the analytical model of
Section~\ref{sec:analytic_voids}.  We represented in the left panel of
Fig.~\ref{fig:ellipticity} the ellipticity distribution of all
observable voids as defined in Section~\ref{sec:definition}. 
The black points give the ellipticity distribution for voids in
the reconstructed displacement field. We estimated error bars assuming a
Poisson distribution of the number of voids in each bin. The red line
is obtained using the method of Appendix~\ref{app:void_tidal}. The
dashed blue line is obtained through the formula of \cite{ParkLee06}, where
no constraints are put on the curvature of the density field where the
ellipticity of the void is measured.

The success of the comparison between the black points and the solid red
curve is striking. It shows the importance of our constraint that we
only look in the negatively curved part of $\sourcePsi$. We note
that this should be a robust feature for voids found with any void
finder. Any probe of the ellipticity in voids looks in regions of the
density field that is strongly underdense, and thus should come mainly
from initially underdense regions. In these primordial underdense
regions, the curvature of the density field is likely to be positive,
which corresponds to a negatively curved $\sourcePsi$ in our
case. {\it Our measured ellipticity distribution in the simulation is very clean
because we rely on features of the displacement field which can be understood in terms
of linear theory even at redshift $z=0$. }

In the right panel of Fig.~\ref{fig:ellipticity}, we show this
same ellipticity distribution but only for ``true voids''. The
comparison between simulation and analytic is also here 
successful. As we noted in Section~\ref{sec:analytics_cosmology}, there
is no real difference between the two models in this case. However,
there is no way a purely geometrical analysis could yield this curve
from the observation of galaxy catalogues, as this requires the
knowledge of the sign of eigenvalues of $T_{l,m}$
(Eq.~\ref{eq:tidal_field}). We note a small shift of $\sim$1-3\%
between the model and the reconstruction. We find, using numerical
experiments with Gaussian random fields, that a fraction of this shift
may be explained by the finite bin size and the very strong steepness
of the distribution represented in this panel.

\subsection{Evolution with redshift}
\label{sec:evolution}

\begin{figure*}
  \includegraphics[width=\hsize]{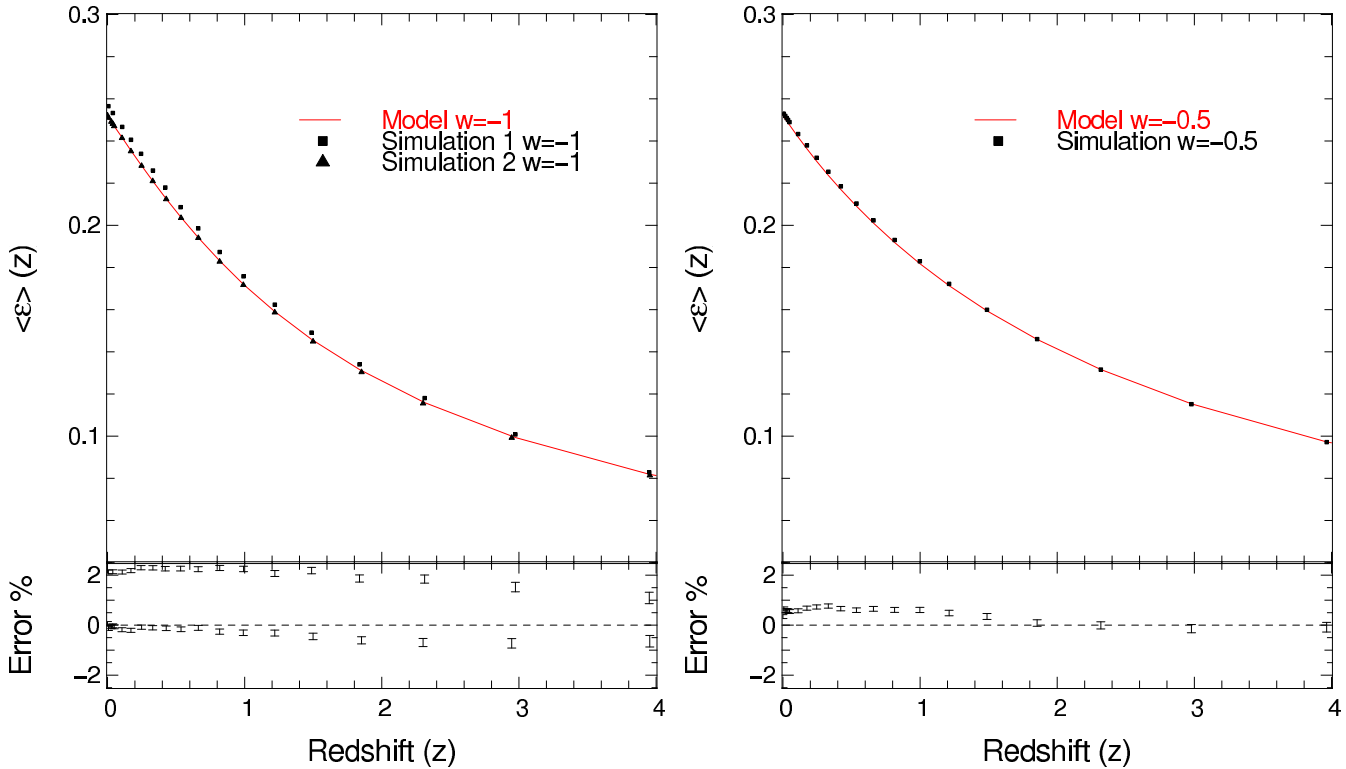}
  \caption{\label{fig:evolution} {\it Evolution of the mean
      ellipticity with redshift} -- We represent the evolution of the
    mean ellipticity with redshift for the two $\Lambda$CDM simulations (left
    panel) and the wCDM simulation (right panel). The mean ellipticities deduced from
    $\Lambda$SIM are represented using square symbols, and the ones
    from $\Lambda$SIM2 using triangular symbols. The solid curve is
    obtained using the statistical model of
    Appendix~\ref{app:void_tidal} and changing $\sigma_8$ according to
    the growth factor as specified in Eq.~\eqref{eq:evolve}. The bottom
    panels give the relative difference, in percentage, between the
    simulation and the analytical model. In the bottom left panel, the
    points at $\sim$2\% correspond to the square symbols}
\end{figure*}

In this section, we focus on the evolution with redshift of the
ellipticity of voids. This evolution has been shown to be analytically
a sensitive probe of $w$ by \cite{LeePark09}. We took snapshots of the simulation at different
redshifts and computed the ellipticity distribution in each of these
snapshots. We note that we must at least have two main differences
compared to what would happen with observations. First, we may have a
systematic effect in the evolution of the mean ellipticity as we are
studying only a single realisation of initial conditions. Second, the
number of available voids should be a non-trivial function of
redshift. Second, because of both volume and selection effects, the error bars should be large at both small and large redshift, while attaining a minimum at some intermediate redshifts
The exact numbers for this second problem depends on the
specific considered galaxy survey, in particular the apparent magnitude limit
and the incompleteness function. 

To compute the predicted ellipticity distribution at any given redshift $z$,
we simply scale the $\sigma_8(z)$ parameter using the growth factor $D(z)$:
\begin{equation}
  \sigma_8(z) = \sigma_8(z=0) \times \frac{D(z)}{D(z=0)}\label{eq:evolve}.
\end{equation}

For clarity we only represent on Fig.~\ref{fig:evolution} the mean ellipticity $\bar{\varepsilon}$, defined as,
\begin{equation}
  \bar{\varepsilon}(z) = \int_0^1 \varepsilon P(\varepsilon|z) \text{d}\varepsilon
\end{equation}
with $P(\varepsilon|z)$ the probability distribution of the
ellipticity $\varepsilon$ at redshift $z$. The red solid line gives
the prediction given by the analytic model of
Section~\ref{sec:analytic_voids}. The black points are obtained from
the simulated field. The error bar on the mean ellipticity is estimated using 
\begin{equation}
  \sigma_{\bar{\varepsilon}} \simeq \frac{\sigma_\varepsilon}{\sqrt{N_\text{voids}}}
\end{equation}
with $\sigma_\varepsilon = 0.02$ as estimated in
Section~\ref{sec:mak_vs_sim} for a smoothing scale of 4\Mpch{}. For
$N_\text{voids}\simeq 10^4$, this gives a typical error of
$\sigma_{\bar{\varepsilon}} = 2\times 10^{-4}$ on the mean. This error bar
corresponds to the uncertainty of the ellipticity derived from the MAK
reconstructed displacement field with respect to the one obtained from
the simulated displacement field. This gives an interval on which the
mean ellipticity can be trusted.

In the left panel of Fig.~\ref{fig:evolution}, we see that the
agreement between the analytical model and the one from the simulated
displacement field (square symbols, ``Simulation 1'') is very good for $w=-1$.
Looking at the relative
error between the model (lower-left panel) and the simulation yields a systematic
$\sim$2\% deviation relative to the expectation. The main reason
is the inexactness of the initial conditions to the finite number of
modes available in the simulation box. Even though the power spectrum
is normalised to $\sigma_8 = 0.77$, the realized $\sigma_8$ of the
displacement field is $0.783$. This produces intrinsically a shift of
1.7\% towards positive errors. It is exactly what we observe.
As we see, this bias is relatively modest. However it should be
observable when we look the small amplitude of the expected
reconstruction errors given by the small error bars.
To check this effect, we rerun another simulation with the exact same
cosmology but with another seed. This time, we measured $\sigma_8=0.7688$
in the initial conditions, this corresponds to a small statistical fluctuation of $-0.15\%$.
We plotted the corresponding evolution of the mean ellipticity in the left
panel of Fig.~\ref{fig:evolution} (triangular symbols, ``Simulation 2'').

This will not prevent applying this method to observations
for two reasons. First, we will marginalise over
the bias and so the systematic shift will disappear.  Second, each
considered slice should be a nearly independent random realisation of a
Gaussian random field normalised to the same $\sigma_8$. Thus the
points should be scattered according to our dashed horizontal line ``0\%'' and
not systematically pushed up or down. 

\subsection{$w=-0.5$ vs $w=-1.0$}
\label{sec:two_cosmologies}

\begin{figure}
  \includegraphics[width=\hsize]{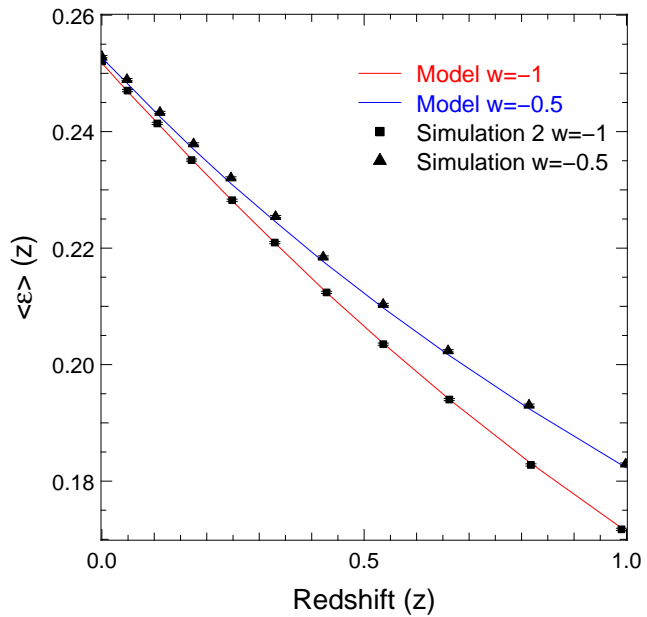}
  \caption{\label{fig:different_cosmo} {\it Difference between w=-1
      and w=-0.5} -- We plotted here the evolution of the mean
    ellipticity with redshift. We used the simulation $\Lambda$SIM2
    (square) and wSIM (triangle). The red solid line gives
    the prediction for $\sigma_8=0.7688$, $w=-1$. The blue solid line
    gives the prediction $\sigma_8=0.773$ and $w=-0.5$. These two
    values of $\sigma_8$ have been computed using the initial
    conditions of the two simulations.  }
\end{figure}

In all the previous sections, we studied the case of a standard
\LCDM{} cosmology with $w=-1$. However, we first started to look at
voids to check if they may be good tracers of the properties of the
dark energy, and in particular of the growth factor. We now focus on
the results obtained from wSIM, a wCDM simulation with
$w=-0.5$ as we specified in Section~\ref{sec:nbody_sample}. The
results are presented in the right panel of Fig.~\ref{fig:evolution}
and in Fig.~\ref{fig:different_cosmo}. We computed that our particular
realisation of the initial conditions had $\sigma_8=0.773$, which is
$0.33\%$ above 0.77. We again note that the discrepancy in the lower-right
panel in Fig.~\ref{fig:different_cosmo} has the
correct systematic shift at low redshift. Taking into account this
shift, as previously, the analytical model seems to follow the results
given by simulation at the $0.1\%$ level, taking into account the
statistical uncertainty. The points obtained from the simulation are
in excellent agreement with the simulation.

Current redshift galaxy catalogues map the Universe at intermediate
redshifts $0 \la z \la 1$. We check if our method is sufficiently
sensitive to distinguish  a $w=-0.5$ from a $w=-1$ cosmology in
Figure~\ref{fig:different_cosmo}. In this figure, we used the
$\sigma_8$ as measured in the simulations to compute the analytical
predictions (red and blue solid curves). We note that even at $z\simeq
0.2$, the behaviour of the two curves is already significantly
different and above statistical uncertainties. If we consider the
whole interval between $z=0$ and $z=1$, the difference is very
significant compared to the uncertainties, with an ellipticity that
changes by $\simeq $35\%.

In all the above we considered catalogues with an important observable
number of voids (typically $\sim$10,000). We do not expect to have
such a high number available in catalogues. Now we try to make an
estimate of the error bars on the mean ellipticity that we may
expect. The SDSS covers one fifth of the sky. We now limit the survey
at $z=0.1$ ($\sim$300\Mpch), and we take a Lagrangian smoothing scale
of 5\Mpch{}. This smoothing scale is motivated by the average density
of galaxies in the SDSS, which in band $r$ is about $1-5\times 10^{-2}
h^3\text{Mpc}^3$ \citep{Blanton03}. This gives a mean separation of
$\sim 2-4$\Mpch{}.
 The equation \eqref{eq:num_maxima} predicts that we should
observe $\sim$1,000 of our voids in this volume when smoothing the
density field in Lagrangian coordinates with a Gaussian of radius
5\Mpch{}, taking into account the survey coverage. If we go to $z=0.2$,
this number should increase to $\sim$9,000. This means that the error
bars should only be moderately larger than the one that we considered
in this work. Typically we expect about three times larger. Even, with
this amount of uncertainty, the comparison to the analytic model
should be able to highly constrain the equation of state of dark
energy at $z \la 0.2$. The Fisher matrix analysis is done in a
companion paper.

\section{Comparison of DIVA to earlier void finders}
\label{sec:discuss_definitions}

In this section, we discuss how our technique is related to the other existing void finders. We try to make a qualitative assessment of its strengths and weaknesses compared to the three classes of void finders define in Section~\ref{sec:intro}.

The void finders of the first class try to find emptier regions in a
distribution of points, which in actual catalogues correspond to galaxies.
The void finder developed by \cite{ElAd97}, and one of its later versions by
\cite{HoyleVogeley2002}, is popularly used in observations
\citep{HV2004,HRV2005,TK06,FosterNelson09}. In these void finders, the
first step consists in classifying galaxies in two types. Galaxies
may lie in a strongly overdense region, in this case we consider it as
a ``wall galaxy''. The other possibility is that they lie in a mildly
underdense region, and they are then called ``field galaxies''. The
exact separation between ``wall galaxies'' and ``field galaxies''
depends on an ad-hoc parameter. This parameter specifies how the local
density of galaxy control the type (field or wall) of the galaxy. The
voids are then grown from regions empty of wall galaxies. 
This classification thus gives a non-trivial dependence of the void sizes and shapes on the galaxy bias and catalogue selection function.
Additionally, this
definition depends on an ad-hoc empirical factor. These issues
make the quantitative study of the geometry of voids difficult, while they find
 voids that correspond to the visual impression given by large scale structures in redshift catalogues of galaxies.

Void finders belonging to the second class look for
particularities in the continuous three dimensional distribution 
of the dark matter traced by galaxies. Of course, from observational data,
one has then to first project and then
smooth the distribution of galaxies to obtain this
distribution. Different techniques are used:
\begin{itemize}

\item[-] One technique is to
adaptively smooth the galaxy distribution either using an SPH
technique \citep[see e.g. ][]{Colombi07} or a Delaunay Tesselation Field
Estimator \citep{dtfe}. Voids must then be identified from the smooth distribution derived using either of
these techniques. One option is to use a scheme similar to
\cite{ElAd97} or \cite{HoyleVogeley2002} to identify shapes of underdensity in
the vicinity of a minima of the density field \citep{Colberg05}. As with void
finders of the first class, the galaxy bias does not affect the positions of the voids but their overall properties. A
second option is to use a Watershed Transform \citep{wvf} to identify
the cavities of the voids. In this case, the galaxy bias does not affect the
structure of the cavities. However, devising an efficient way of relating
the geometrical properties of these cavities to the cosmology, which corresponds to studying Morse theory, could well
be non-trivial \citep{Jost08}. Some work to study the skeleton (also called the ``cosmic spine'') of Large Scale structures in this theory
has been recently done by \cite{AragonCalvo08}, \cite{Pogosyan08} and \cite{Sousbie09}.

\item[-] A second technique consists in using the
local density estimated using the Voronoi diagram of a Delaunay
tessellation to locate minima \citep{zobov}. The particles are first
grouped in zones. Each particle is assigned to a zone on to
where this particle is attracted if it followed the density
gradient as in the watershed technique. Each zone is defined to be a void.
But it is also possible
to define a hierarchy of voids by checking, for two neighbouring
voids, which of the two has the lowest density at the minima.
The zones are then grouped and a probability of being a void is assigned
depending on the contrast between the density at the ridge of the void and its depth. 
\end{itemize}
This void finder has the advantage of defining voids in terms of
topology of the density field, which is easier to handle from a
theoretical point of view and may better define a void in terms of
dark matter. Still, we are faced with the task of relating the shape of the
voids that are found by these algorithms, which is non-local by nature, to
cosmology. As mentioned previously for void finders of the first class, this seems to be non-trivial.

Void finders of the third class use the inferred dark matter
distribution but they do a dynamical analysis to infer the
location of these voids.  DIVA belongs to these class of void finders.
There are two advantages of looking at dynamics for voids. (i)
It gives a much more physical and intuitive definition of these
structures: voids corresponds to places in the universe from which the
matter {\it is really escaping} and not gravitationally unstable at present time. (ii) Using this criterion, one is
bound to use either the velocity field or the displacement
field. These two quantities are highly linear. It has been directly shown for velocity fields by \cite{CCK03} and indirectly shown by \cite{moh2005} and \cite{lavaux08} for the displacement field. This linearity 
helps us at constructing an analytical statistical model of the voids.
\cite{Hahn07} and \cite{VoidsGravity08} attempted to classify 
structures according to a criterion on the gravitational field.  However we
may highlight two very important differences compared to the approach we are following here:
\begin{itemize}
\item[-] we are using a purely Lagrangian method and it takes into account the
  true evolution of the void and not how virtual tracers in the void should move now,
\item[-] we put an exact natural threshold on eigenvalues to classify
  voids. This is in contrast with \cite{VoidsGravity08} where they need to
  to put a threshold on eigenvalues depending on an estimated collapse time.
  In our case, everything is already integrated in the definition of 
  the displacement field.
\end{itemize}
Moreover, the Monge-Amp\`ere-Kantorovitch reconstruction presents the
two advantages of: (i) never diverging in the neighbourhood of large
density fluctuations, compared to a pure gravitational approach, (ii)
recovering exactly the Zel'dovich approximation in the neighbourhood
of centres of voids.

As for the other void finders,
DIVA depends on galaxy bias. We recall that the bias $b$ is defined with
\begin{equation}
  \delta_g \simeq b \delta_m ,
\end{equation}
with $\delta_g$ the density fluctuations of galaxies and $\delta_m$ the density fluctuations of matter. As MAK is essentially reconstructing the Zel'dovich displacement 
in underdense regions, and that the Zel'dovich potential is proportional to
density fluctuations, the MAK displacement should also be mostly linear with the bias.

We describe in a
second paper (Lavaux \& Wandelt 2009, in prep.) how to relate the
volume of the voids that we find in Lagrangian coordinates to the voids that we
observe in Eulerian coordinates.


\section{Conclusion}
\label{sec:conclusion}

We have described a new technique to identify and characterise voids
in Large Scale structures. Using the MAK reconstruction, we have been
able to define void centres in Lagrangian coordinates by assimilated them to the maxima of
the divergence $\sourcePsi$ of the displacement field, interpreted as its
source term. The scalar field $\sourcePsi$ has the
interesting property of being nearly equal to the opposite of the linearly
extrapolated primordial density field \citep{moh2005}. This allowed us
to consider that the statistics of those two fields were equal. Using
that, we made predictions on the number of voids
in Lagrangian coordinates, along with their ellipticities defined
using the eigenvalues of the curvature of $\sourcePsi$. 

We tested our model using $N$-body
simulations with different cosmologies ($w=-1$ and $w=-0.5$). We
checked, using the largest Lagrangian reconstruction run so far, that
MAK is capable of recovering the ellipticity of individual voids to the
order of a few percent. We highlighted the importance of using our
model for the statistics of the eigenvalues of the curvature
instead of the formula of
\cite{Dor70} for the particular case of voids. We showed that our
analytical model agrees within $\sim$0.1-0.3\% to
results obtained with MAK and the displacement field obtained from the
simulation. We expect our method to be able to provide a very
promising constraint on the equation of state of the Dark Energy of
the late universe, especially given the notable accuracy of the prediction
that we obtained Fig.~\ref{fig:different_cosmo}.

We intend to pursue this work to continue characterising analytically
the voids found by DIVA, in particular the evolution of the
number of voids and their size distribution.
We will make further robustness tests using mock
catalogues, including especially redshift distortion effects. We also
would like to apply our method to the Luminous Red Galaxy sample of
the SDSS DR7 \citep{SDSSDR7}.

\section*{Acknowledgements}

This research was supported in part by the National Science Foundation
through TeraGrid resources provided by the NCSA. TeraGrid systems are
hosted by Indiana University, LONI, NCAR, NCSA, NICS, ORNL, PSC, 
Purdue University, SDSC, TACC and UC/ANL. 
The authors acknowledge financial support from NSF grant AST 07-08849.
We acknowledge the hospitality of the California Institute of Technology, 
where the authors completed most of this work. 
We thank the referee, Rien van de Weygaert, for his useful comments and 
suggestions.


\appendix
\section{Eigenvalues probability distribution}
\label{app:pdf_lambda}

In this appendix, we derive the unconditional distribution of the
eigenvalues of the Jacobian matrix of the displacement field in the
Zel'dovich regime.  Starting from the potential $\Phi({\bf q})$ of the
displacement field at the comoving Lagrangian coordinate ${\bf q}$, we
define the Jacobian matrix of the displacement field
\begin{equation}
  T_{i,j} = \frac{\partial \Phi}{\partial q_i \partial q_j}\,.
\end{equation}
This matrix is real and symmetric. We assume the components
to be normally distributed and to be as
given by \citep[][Appendix A]{BBKS}:
\begin{equation}
  \langle T_{i,j} T_{k,l} \rangle = \frac{\sigma^2}{15} \left(\delta_{i,j}\delta_{k,l} + \delta_{i,k}\delta_{j,l} + \delta_{i,l}\delta_{j,k}\right)\,,
\end{equation}
with $\sigma$ the standard deviation of the density field.
As it is a $3 \times 3$ real symmetric matrix, there are only 6 independent
components.  We label these components with $A=1\ldots 6$, where each
number refer to the $(i,j)$ couples $(1,1),(2,2),(3,3),(1,2),(1,3)$
and $(2,3)$. We may now write the matrix $V$ of the variances
\begin{equation}
  V = \frac{\sigma^2}{15}
        \left(
            \begin{array}{cccccc}
              3 & 1 & 1 & 0 & 0 & 0 \\
              1 & 3 & 1 & 0 & 0 & 0 \\
              1 & 1 & 3 & 0 & 0 & 0 \\
              0 & 0 & 0 & 1 & 0 & 0 \\
              0 & 0 & 0 & 0 & 1 & 0 \\
              0 & 0 & 0 & 0 & 0 & 1 \\             
            \end{array}
        \right)
\end{equation}
thus the covariance matrix $C$ is
\begin{equation}
  C = V^{-1} = \frac{15}{10\sigma^2}
                \left(
                  \begin{array}{cccccc}
                    4 & -1 & -1 & 0 & 0 & 0 \\
                    -1 & 4 & -1 & 0 & 0 & 0 \\
                    -1 & -1 & 4 & 0 & 0 & 0 \\
                     0 & 0 & 0 & 10 & 0 & 0 \\
                     0 & 0 & 0 & 0 & 10 & 0 \\
                     0 & 0 & 0 & 0 & 0 & 10 \\             
                  \end{array}
                  \right)
\end{equation}
The unconditional probability $\text{d}P(\{T_A\})$ of observing these
components at a point in the universe is given by
\begin{equation}
  \text{d}P(\{T_A\}|\sigma) = \frac{3375}{16 \sqrt{5} \sigma^6 \pi^3} \text{exp}\left( - \frac{1}{2} \sum_{A,B=1}^6 C_{A,B} T_A T_B \right) \prod_{A=1}^6 \text{d}T_A\label{eq:fullGaussProba}
\end{equation}
We want now to make a change of variables and get the probability
distribution of the eigenvalues of $T_{i,j}$. The appendix B of
\cite{BBKS} shows that the infinitesimal volume is transformed
according to:
\begin{eqnarray}
  \prod_{A=1}^6 \text{d}T_A = \left|(\lambda_1-\lambda_2)(\lambda_1-\lambda_3)(\lambda_2-\lambda_3)\right| \text{d}\lambda_1\text{d}\lambda_2\text{d}\lambda_3\text{d}\Omega_{S^3}\,,
\end{eqnarray}
with $\text{d}\Omega_{S^3}$ the infinitesimal rotation of the
hypersphere of dimension 3.  The quadratic form $Q$ in the argument of
the exponential of Eq.~\eqref{eq:fullGaussProba} may be expanded
\begin{multline}
  Q = \sum_{A,B=1}^6 C_{A,B} T_A T_B = \frac{15}{\sigma^2}\left(4 (T_1^2 + T_2^2 + T_3^2) \right. \\
  \left. - 2 (T_1 T_2 + T_1 T_3 + T_2 T_3) + T_4^2 + T_5^2 + T_6^2\right)\,.
\end{multline}
Linear algebra tells us that the scalar quantities
\begin{equation}
  K_1 = T_1 + T_2 + T_3 \text{ and } K_2 = T_1 T_2 + T_2 T_3 + T_1 T_3 - T_4^2 - T_5^2 - T_6^2
\end{equation}
are invariant by change of vector basis. 
The quadratic form may then be rewritten:
\begin{equation}
  Q = \frac{6}{\sigma^2}\left(K_1^2 - \frac{5}{2} K_2\right)
\end{equation}
We may now express $K_1$ and $K_2$ in terms of the eigenvalues:
\begin{eqnarray}
  K_1 & = & \lambda_1 + \lambda_2 + \lambda_3\,, \\
  K_2 & = & \lambda_1 \lambda_3 + \lambda_1 \lambda_2 + \lambda_2 \lambda_3 \,.
\end{eqnarray}
After integrating the expression \eqref{eq:fullGaussProba} over $S^3$,
we obtain the unconditional probability that the Jacobian matrix has
the three ordered eigenvalues $\lambda_1$, $\lambda_2$ and
$\lambda_3$:
\begin{multline}
  P(\lambda_1,\lambda_2,\lambda_3|\sigma)  =\\
   \frac{3375}{8\sqrt{5}\sigma^6 \pi} \text{exp}\left[\frac{3\left(2 K_1^2 - 5 K_2\right)}{2\sigma^2}\right] |(\lambda_1-\lambda_2)(\lambda_1-\lambda_3)(\lambda_2-\lambda_3)|.
\end{multline}
This equation corresponds to the distribution derived by \cite{Dor70}. 


\section{Probability distribution of tidal field in voids}
\label{app:void_tidal}

We show in this Appendix how to compute the form of the probability
distribution of the gravitational tidal field in the case of voids. 
The technique involved in this derivation looks much like the one used
in Appendix~\ref{app:pdf_lambda} but with an extra complication due to
correlations with the density field.
By definition, void centres are chosen to be maxima of 
\begin{equation}
  \sourcePsi({\bf q}) = \sum_{i=1}^3 \frac{\partial^2 \Phi}{\partial q_i^2}\,.
\end{equation}
with $\Phi$ defined as in \S~\ref{sec:mak}, the scalar potential of
the displacement field. We will assume that, for voids, we are fully
in the Zel'dovich regime and thus we can equate $\Phi$ as given by MAK
to the potential of the Zel'dovich displacement at high redshift. In
this case, $\sourcePsi$ corresponds the primordial density
fluctuations scaled linearly to $z=0$. 

Being a maxima of $\sourcePsi$ means the displacement field satisfies two
properties: the gradient of $\sourcePsi$ is null and the Hessian matrix
\begin{equation}
  H_{i,j} = \frac{\partial \sourcePsi}{\partial q_i \partial q_j}
\end{equation}
of $\sourcePsi$ is negative-definite. Thus our aim is to compute the
probability of the Jacobian matrix of the displacement field given
that $H_{i,j}$ is symmetric negative-definite. We write this
probability $P(T_{i,j} | H_{l,m} < 0)$.

We assume that the gravitational potential is a Gaussian random field determined by the
power spectrum of matter density fluctuation $P(k)$, with $k$ a wave number. 
We compute the correlation between those two fields:
\begin{eqnarray}
  \langle T_{i,j} T_{l,m} \rangle  & = & \frac{\sigma^2_T}{3} \left(\delta_{i,j}\delta_{l,m} + \delta_{i,l}\delta_{j,m} + \delta_{i,m}\delta_{j,l}\right)\,, \\[.5cm]
  \langle H_{i,j} H_{l,m} \rangle & = & \frac{\sigma^2_H}{3} \left(\delta_{i,j}\delta_{l,m} + \delta_{i,l}\delta_{j,m} + \delta_{i,m}\delta_{j,l}\right)\,, \\[.5cm]
  \langle T_{i,j} H_{l,m} \rangle & = & \frac{\Gamma_{TH}}{3} \left(\delta_{i,j}\delta_{l,m} + \delta_{i,l}\delta_{j,m} + \delta_{i,m}\delta_{j,l}\right)\,,
\end{eqnarray}
with
\begin{equation}
  \sigma^2_T = \mathcal{S}_2 = \sigma^2 \text{ and } \sigma^2_H = \mathcal{S}_{6}\text{ and } \Gamma_{TH} = - \mathcal{S}_4,
\end{equation}
$\sigma^2$ the variance of the density field and
\begin{equation}
  \mathcal{S}_n = \frac{1}{10 \pi^2} \int_{k=0}^{+\infty} k^n P(k)\text{d}k\,.
\end{equation}
To reduce the complexity of the correlation matrix, we now use the reduced random variables defined as follow
\begin{eqnarray}
  \tilde{T}_{i,j} & = & \frac{T_{i,j}}{\sigma^2_T} \\
  \tilde{H}_{i,j} & = & \frac{H_{i,j}}{\sigma^2_H}\,.
\end{eqnarray}
and the reduced correlation
\begin{equation}
  r = \frac{\Gamma_{TH}}{\sigma_T \sigma_H}\label{eq:correl_curvature}.
\end{equation}

As $T$ and $H$ are $3\times 3$ real symmetric matrices, there are only
6 independent components for each of these matrices. As in the Appendix~\ref{app:pdf_lambda}, we label these
components with $A=1\ldots 6$, where each number refer to the $(i,j)$
couples $(1,1),(2,2),(3,3),(1,2),(1,3)$ and $(2,3)$. The matrix $V$ of
the variance of the 12 reduced components may be formally written,
using a block-matrix representation:
\begin{equation}
  V = \left( \begin{array}{cc}
                  A & r A \\
                  r A & A
  \end{array} \right)
\end{equation}
and
\begin{equation}
  A =  \left( \begin{array}{cccccc}
      1 & \frac{1}{3} & \frac{1}{3} & 0 & 0 & 0 \\
      \frac{1}{3} & 1 & \frac{1}{3} & 0 & 0 & 0 \\
      \frac{1}{3} & \frac{1}{3} & 1 & 0 & 0 & 0 \\
      0 & 0 & 0 & 1 & 0 & 0  \\
      0 & 0 & 0 & 0 & 1 & 0  \\
      0 & 0 & 0 & 0 & 0 & 1  
      \end{array}\right).
\end{equation}
The inverse may be computed straightforwardly
\begin{equation}
  C = M^{-1} = \frac{1}{1-r^2}\left(\begin{array}{cc}
                 A^{-1} & -r A^{-1} \\
                 -r A^{-1} & A^{-1}
                 \end{array}\right)
\end{equation}

Now, we may express the joint probability $P(T,H)$ of observing a
tidal field $T$ for the gravitational potential and a curvature $H$
for the density field using the covariance matrix $C=M^{-1}$.
\begin{equation}
  P(T,H|r) = \frac{\sqrt{|\det C|}}{(2\pi)^{6}}\text{exp} \left( -\frac{1}{2}{}^t Y C Y\right)
\end{equation}
with $Y = (T,H)$. Now the probability of observing some matrix $T$ given that $H$ must be positive could be computed formally:
\begin{equation}
  P(T|r,H<0) = \int_{H < 0} P(T,H|r)\text{d}H \label{eq:formal_positivity}
\end{equation}
However, it is quite involved to find an analytic expression of this
integral in function of the eigenvalues of $T$. We propose to sample
this distribution instead of computing this integral.

One can prove that the conditional probability
$P(T|H)$ may be written
\begin{multline}
  P(T|H,r) = \frac{P(T,H|r)}{\int_{T} P(T,H|r)} = \\
  \frac{\sqrt{|\det A|}}{(2\pi)^3}\text{exp}\left(-\frac{1}{2} {}^t(T - r H) A (T - r H)\right)\label{eq:conditional_T}
\end{multline}
Thus, Equation~\eqref{eq:formal_positivity} may be re-expressed as:
\begin{equation}
  P(T|r,H<0) = \int_{H < 0} P(T|r,H) P(H)\text{d}H
\end{equation}
with $P(H)$ the probability of getting a random symmetric matrix $H$,
with the covariance matrix $A$. The method is now the following:
\begin{itemize}
  \item we generate a random symmetric matrix $H$, if it is negative, we accept
    it, in the other case we try again;
  \item we generate a matrix $T$, following the probability given by
    Equation~\eqref{eq:conditional_T};
  \item we compute and store the eigenvalue of this matrix $T$, after
    multiplication by $\sigma_T$.
\end{itemize}
That way the joint probability distribution
$P(\lambda_1,\lambda_2,\lambda_3|\sigma_T,r,H<0)$ is correctly sampled, even
though we do not have any explicit expression of it.

\section{Local tidal ellipticity vs. global volume ellipticity}
\label{app:local_vs_global_ellipticity}

In this appendix, we try to relate the two ellipticities $\varepsilon_{vol}$ and $\varepsilon_\text{DIVA}$ defined in Eq.\eqref{eq:epsilon} and \eqref{eq:epsilon_shape}. To do that, we will make use of Zel'dovich approximation in voids, which has been shown to be a relatively precise modeling of the void evolution.
The inertial mass tensor writes as:
\begin{equation}
  M = a^2 \mathcal{I} - K
\end{equation}
with $\mathcal{I}$ the $3\times 3$ identity matrix,
\begin{eqnarray}
   a^2 & = & \int_{\mathcal{V}} \text{d}^3{\bf q}\, ||{\bf x}({\bf q})-\bar{\bf x}||^2 \\
   K_{i,j} & = &  \int_{\mathcal{V}} \text{d}^3{\bf q}\,(x_i({\bf q}) - \bar{x}_i) (x_j({\bf q}) - \bar{x}_j),
\end{eqnarray}
with ${\bf q}$ the Lagrangian coordinates, $\mathcal{V}$ the Lagrangian domain of the considered void, $\bar{\bf x}$ the centre of mass of the $\mathcal{V}$ in Eulerian coordinates. With this parametrization, the volume ellipticity $\varepsilon_\text{vol}$ simplifies as
\begin{equation}
  \varepsilon_\text{vol} = 1 - \left(\frac{J_1}{J_3}\right)^{1/4}
\end{equation}
with $J_1$ and $J_3$ the smallest and largest eigenvalues of $K$.
We may write exactly:
\begin{equation}
   {\bf x}({\bf q}) = {\bf q} + \bm{\Psi}({\bf q}).
\end{equation}
We now expand $\bm{\Psi}$ to first order around the position of the centre of mass in Lagrangian coordinates $\bar{\bf q}$
\begin{equation}
  \Psi_i({\bf q}) = \Psi_i(\bar{\bf q}) + \frac{\partial \Psi_i}{\partial q_j}(q_j - \bar{q}_j).
\end{equation}
Using Zel'dovich approximation we identify ${\partial \Psi_i}/{\partial q_j}$ and $T_{i,j}$ given in Eq.~\eqref{eq:tidal_def}. We now reexpress $K_{i,j}$
\begin{eqnarray}
  K_{i,j}  &  = & L_{i,j} + T_{i,k} L_{k,j} + T_{j,k} L_{k,i} \\
  L_{i,j} & = & \int_{\mathcal{V}} \text{d}^3{\bf q}\, (q_i - \bar{q}_i)(q_j-\bar{q}_j). 
\end{eqnarray}
As voids, in Lagrangian coordinates, should be mostly isotropic the Lagrangian inertial tensor $L_{i,j}$ must be diagonal: 
\begin{equation}
  L_{i,j} = \frac{1}{3} a_\text{L}^2 \delta_{i,j},
\end{equation}
with
\begin{equation}
  a^2_\text{L} = \int_\mathcal{V} \text{d}^3{\bf q} ||{\bf q}-\bar{\bf q}||^2.
\end{equation}
This assumption is verified in average by linear theory but may be broken for some specific voids.
In the case where voids are effectively isotropic, the inertial mass tensor $K_{i,j}$ is extremely simplified:
\begin{equation}
  K_{i,j}  = \frac{a_\text{L}^2}{3} \left(\delta_{i,j} + 2 T_{i,j}\right).
\end{equation}
The eigenvalues of $K$ are thus
\begin{equation}
  J_i \propto 1 + 2 \lambda_i \simeq (1 + \lambda_i)^2.
\end{equation}
The volume ellipticity may thus be related to the tidal ellipticity as
\begin{equation}
  \varepsilon_\text{vol} = 1 - \left(\frac{J_1}{J_3}\right)^{1/4} \simeq 1 - \left(\frac{1 + \lambda_1}{1 + \lambda_3}\right)^{1/2} = \varepsilon_\text{DIVA}
\end{equation}

\end{document}